\documentclass[useAMS]{mn2e}

\usepackage{amssymb,amsmath,psfig,times,url}
\def\gsim{ \lower .75ex \hbox{$\sim$} \llap{\raise .27ex \hbox{$>$}} }
\def\lsim{ \lower .75ex\hbox{$\sim$} \llap{\raise .27ex \hbox{$<$}} }

\def\ergcms{{\rm\thinspace erg \thinspace cm^{-2} \thinspace s^{-1}}}
\def\ergs{{\rm\thinspace erg \thinspace s^{-1}}}

\def\sc{Schwarzschild}
\def\beq{\begin{equation}}
\def\eeq{\end{equation}}

\def\sc{Schwarzschild}

\title[Blazar candidates at $z>4$]
{Blazar candidates beyond redshift 4 observed with GROND}
\author[T. Sbarrato et al.]
{T. Sbarrato$^{1,2}$\thanks{Email: tullia.sbarrato@brera.inaf.it}, 
G. Ghisellini$^2$, 
M. Nardini$^3$,
G. Tagliaferri$^2$, 
J. Greiner$^4$, 
A. Rau$^4$, 
P. Schady$^4$
\\
$^1$Dipartimento di Scienza e Alta Tecnologia, Universit\`a dell'Insubria, 
Via Valleggio 11, I--22100 Como, Italy\\
$^2$INAF -- Osservatorio Astronomico di Brera, Via Bianchi 46, I--23807 Merate, Italy\\
$^3$Dipartimento di Fisica G. Occhialini, Universit\`a di Milano Bicocca, 
Piazza della Scienza 3, I--20126 Milano, Italy\\
$^4$Max--Planck--Institut f\"ur extraterrestrische Physik, Giessenbachstrasse 1, 85748 Garching, Germany \\
}
\date{Accepted 2013 May 16.  Received 2013 May 9; in original form 2013 March 27}

\begin{document}  

\maketitle

\begin{abstract}
The search for extremely massive high redshift blazars is essential to put strong 
constraints on the supermassive black holes formation theories.
Up to now, the few blazars known to have a redshift larger than 4
have been discovered serendipitously.
We try a more systematic approach.
Assuming radio--loudness as a proxy for the jet orientation, 
we select a sample of extremely radio--loud quasars.
We measure their black hole masses with a method based on fitting 
the thermal emission from the accretion disc.
We achieve a precision of a factor of two for our measures, 
thanks to the observations 
performed with the Gamma--Ray Burst Optical Near--Infrared Detector (GROND).
The infrared to optical GROND data allow us to observe directly the peak of the disc emission, thus 
constraining the overall disc luminosity.
We obtain a small range of masses, that peaks at $10^{9.3}M_\odot$. 
If some of our candidates will be confirmed as blazars, these results would introduce  
interesting constraints on the mass function of extremely massive black holes 
at very high redshift.
Moreover, all our blazar candidates have high accretion rates. 
This result, along with the high masses, opens an interesting view on the
need of a fast growth of the heaviest black holes at very high redshift.
\end{abstract}


\section{Introduction}

\begin{figure}
\vskip -0.6 cm 
\hskip -0.1 cm
\psfig{figure=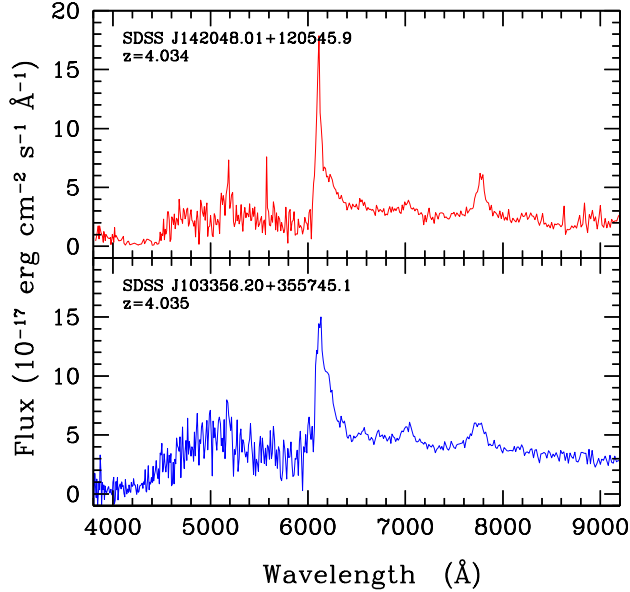,width=9cm,height=9cm}
\vskip -0.5 cm 
\caption{
Comparison of the optical spectra of an extremely radio--loud 
(upper panel) and a radio--quiet AGN (lower panel) with the 
same redshift, both from the sample of S11. 
Note that there are no differences between the two spectra that 
could suggest their difference in radio--loudness.
}
\label{spec}
\end{figure}

Blazars are radio--loud Active Galactic Nuclei (AGN) with relativistic 
jets directed at a small angle from our line--of--sight.
It is convenient to define blazars those sources whose jet makes an
angle $\theta_{\rm v}$ to our line of sight comparable to or 
smaller than $1/\Gamma$, 
where $\Gamma$ is the bulk Lorentz factor of the emitting region of the jet.
At these angles the jet emission is strongly boosted by relativistic effects.
Hence, blazars should be in principle visible even at very high redshift.
Nevertheless, up to now, only few of them are known at $z>4$, 
and they have all been discovered serendipitously.
The typical blazar Spectral Energy Distribution (SED) is dominated by the non--thermal jet 
emission, and is characterized by two non--thermal humps, respectively attributed to 
synchrotron (at low frequency) and inverse Compton (high frequency) processes.
Generally, the emission from the other structures of the blazar 
is not or barely visible, because of this jet dominance 
(i.e.\ the thermal emission from the accretion disc, that one should expect to observe 
in the optical--UV band, is totally or partially covered 
by the synchrotron component). 
As the bolometric luminosity increases, 
the two humps shift to lower frequencies 
(for a different interpretation of this effect, see Giommi et al.\ 2012).
Indeed, the synchrotron component can shift enough to let  
the thermal emission from the accretion disc to emerge. 
In this case, the blazar would show 
to the observer an optical spectrum with features completely 
similar to those displayed by a radio--quiet AGN with an 
analogous accretion disc, not contaminated by other components.
To illustrate this point, in Fig. \ref{spec} we show 
the optical spectra of two quasars at the same redshift:
one is radio--quiet, and the other is strongly radio--loud. 
The two spectra are indistinguishable.
This is because at 
% , that peaks in the optical--NIR range.
% This is what happens to powerful high--redshift blazars.
$z>4$ we can only see the most powerful objects, and 
there is a correlation between the luminosity
of the accretion disc, the jet power and the 
peak frequencies of the non--thermal emission 
(Ghisellini et al.\ 2011, Sbarrato et al.\ 2012a, Ghisellini et al.\ 2010a,b). 
For larger luminosities, the synchrotron peak
shifts to sub--mm 
frequencies, leaving the accretion disc component
``naked" and thus observable.
An analogous effect occurs for the high--energy hump, 
whose peak is shifted below 100 MeV,
%shifting its peak below 100 MeV,
%In these high--redshift and high--luminosity objects, in fact,
%the inverse Compton component peaks  below 100 MeV, 
making its detection 
% 5 through $\gamma$--ray telescopes, like 
by the Large Area Telescope (LAT) onboard the 
{\it Fermi Gamma--Ray Space Telescope} (Atwood et al., 2009) unlikely.
Instead, a powerful blazar can be more easily detected 
at frequencies just before the inverse Compton peak 
by hard X--ray instruments, e.g.\ the Burst Alert Telescope (BAT) onboard 
the {\it Swift} satellite (Gehrels et al.\ 2004).
This effect is enhanced by the high redshift itself, therefore it is surely 
more effective to use a hard X--ray telescope instead of a $\gamma$--ray 
instrument to detect high power $z>4$ blazars.

By measuring the power of the source close to where it peaks, 
X--ray data are also crucial to classify the source as a blazar,
since misaligned or even slightly misaligned jets correspond to
much less luminous sources than their aligned counterparts.
Following these guidelines, we could recently classify B2 1023+25, 
a $z\sim$5.3 radio--loud source, as a blazar, thank to its hard and strong
X--ray emission (Sbarrato et al. 2012b).

%Lower redshift blazars are generally identified through their $\gamma$--ray emission,  
%since it indicates a prominent inverse Compton hump.
%In case those data are not available, the only way to confirm the presence of this 
%high--energy component is to observe a hard and intense X--ray flux, as in the case 
%of B2 1023+25, recently identified as a blazar in our previous work (Sbarrato et al.\ 2012).

The importance of observing blazars at $z>4$ resides in the fact
that we expect their black holes (BH) to be very massive, 
since we observe only the most powerful objects:  
$M_{\rm BH}$ can even exceed a billion solar masses.

A large number of extremely massive BHs at $z>4$ could potentially put strong 
constraints on the study of their formation.
Because of their peculiar orientation, they have a great statistical relevance:
for each observed blazar, in fact, there should be $2\Gamma^2= 450(\Gamma/15)^2$ 
analogous radio--loud AGN with their jets directed in random directions 
(Ghisellini et al.\ 2010a; Volonteri et al.\ 2011)
Discovering even a few blazars at high--$z$ has a big impact
on the study of early BHs, such as their comoving density at high--redshift.
(Volonteri et al.\ 2011; Rau et al.\ 2012).
 
For these reasons, the search for high--$z$ blazars can be as rewarding as 
the search for high--$z$ radio--quiet sources, even if radio--quiet AGN are intrinsically
more numerous than radio--loud ones.
But to efficiently pursue this goal, we must search blazars in a systematic way, instead 
of serendipitously as happened up to now.
A systematic approach is possible if a complete quasar catalog is used 
as a starting point for searching for blazar candidates.
In this work, we use the Quasar Catalog (Schneider et al.\ 2010)
obtained from the $7^{\rm th}$ Data Release (DR7) 
of the Sloan Digital Sky Survey (SDSS, York et al. 2000).
% Indeed, the optical features of high--$z$ powerful radio--loud AGN (including blazars)
% are completely indistinguishable from the 
% features of radio--quiet AGN, as can be seen in Figure \ref{spec}.
% In the two panels, the spectra of an extremely radio--loud (upper) 
% and a radio--quiet (lower) AGN at the same redshift are shown.
% No hints that can suggest such a deep difference 
% in their radio--loudness is present.
% Hence, it is clear that chasing blazar candidates among a general 
% large sample of quasars (even dominated by radio--quiet objects) 
% is an efficient approach. 
% Moreover, we will show in our work that 
% All the features related to the accretion disc have no clear differences 
% in the case of a radio--quiet or a radio--loud AGN.

%----------------------------------------------------
%----------------------------------------------------

Starting from this 
well defined (namely, flux limited and covering a known portion of the sky) 
sample, we can select the best blazar candidates among the most radio--loud sources.
In this way, once we will confirm the blazar nature of the candidates, we can construct the number
density not only of the confirmed blazars, but also of their parent population (i.e. 
their misaligned counterparts).
%If, in addition, we can 
Moreover, by measuring the black hole mass in these systems,
we can construct the number density of black holes of a given mass at high redshifts.

% To identify them, we  select a group of possible blazar candidates on the basis of 
% their radio--loudness.
% and black hole mass.
% The $M_{\rm BH}$ is important in order to characterize completely the candidates, 
% principally in the perspective of a BH density study.

In this work we will select a sample of blazar candidates 
chosen among the sample of SDSS quasars of Shen et al. (2011, hereafter S11).
The considered region of the sky is covered by the 
FIRST radio survey
(1.4 GHz, with a flux limit of 1 mJy; White et al.\ 1997).
We will select all quasars in that sample with $z>4$ and large radio--loudness.
For each of them we will construct the IR--optical--UV spectral energy distribution (SED).
All objects visible from La Silla (Chile) have been observed with the 
Gamma--Ray Burst Optical Near--Infrared Detector (GROND; 
Greiner et al. 2008), providing 
photometric data from the near IR to the optical.
All objects have a SDSS spectrum, and all (but 4) are detected in the far IR
by the Wide--field Infrared Survey Explorer (WISE; Wright et al.\ 2010).
With the good wavelength coverage that is obtained, we are able to 
reliably estimate the black hole mass of the objects, by fitting the data with a 
standard accretion disc model.
Along with the mass, we obtain the overall disc luminosity and, hence, the 
Eddington ratio for the whole sample.
This work will provide a complete sample (i.e. flux limited) 
of possible blazar candidates, well described in their thermal emission and 
principal features.
\S \ref{sample} describes the sample selection, \S \ref{grond} 
shows the details of the GROND observations, \S \ref{sed} explains 
the available data and the fitting method, in \S \ref{discussion} we 
discuss the results and their implications, while in \S \ref{concl} 
we summarize our work.

We adopt a flat cosmology with $H_0=70$ km s$^{-1}$ Mpc$^{-1}$ and
$\Omega_{\rm M}=0.3$.

\section{The sample}
\label{sample}

% We need to select in a systematic way a large number of radio--loud AGN at $z>4$.
The SDSS provides the largest publicly available 
Quasar Catalog (Schneider et al.\ 2010), complete with both photometric and spectroscopic data
of 105,783  optically selected quasars.
This catalog has been spectroscopically analyzed by S11, 
who derived continuum and lines features, along with different virial 
mass estimates for all the objects.
S11 cross--correlated the SDSS Quasar Catalog with the FIRST radio catalog,
providing the 1.4 GHz radio fluxes and radio--loudnesses for all 
SDSS AGN with a FIRST radio counterpart.

In order to select good high--redshift blazar candidates, we limit our research 
to the quasars with a radio matching and collect the most radio--loud 
quasars with $z>4$.
We obtain a sample of 31 quasars with $z>4$ and radio--loudness 
\begin{equation}
R \equiv \left[ {F_{\rm 1.4GHz}\over F_{\rm 2500\AA} } \right]_{\rm restframe}>100.
\end{equation}
The limit in radio--loudness should permit to preferentially select the objects with their 
jets oriented closer to our line--of--sight.
This provides us a group of 31 possible candidates listed in 
%This should allow us to narrow down the group of possible candidates.
Tab.\ \ref{31} along with their most illustrative data.
We include from S11 redshift, radio--loudness, radio flux, bolometric and CIV line 
luminosities and the best Black Hole mass estimate, where available.
For those where the CIV line was visible, we include the overall Broad Line Region 
luminosity ($L_{\rm BLR}$), derived following the template by Francis et al.\ (1991).
For the few objects already observed in the X--rays, we report in Tab.\  \ref{31} 
the X--ray spectral index and flux (in the 0.3--10 keV band).

It is indeed worth to notice that among the 31 selected quasar, 
7 have already an X--ray detection.
2 of them can be classified as blazars, thanks to their hard and intense X--ray fluxes 
(i.e.\ SDSS J083946.22+511202.8 and SDSS J151002.92+570243.3).
Specifically, SDSS J151002.92+570243.3 is one of the firsts high redshift 
($z>4$) blazars ever discovered (Moran \& Helfand, 1997).
Moreover, we identified in a previous work our best candidate B2 1023+25 as a blazar 
(Sbarrato et al.\ 2012b), thanks to the {\it Swift} X--ray data (Gehrels et al.\ 2004).

Here, we want to estimate the central black hole mass ($M_{\rm BH}$) of our blazar candidates.
Since at $z>4$ the virial mass method has to rely only on the CIV line, 
we use a different method to estimate $M_{\rm BH}$, based on the direct fit of the accretion disc 
(see \S \ref{mass}). 
In order to fit it properly, we need a good coverage of the NIR--optical wavelength range.
We used the spectroscopic and 
photometric optical data provided by the SDSS DR7, along with the publicly 
available data from WISE\footnote{Data retrieved from \url{http://irsa.ipac.caltech.edu/}}. 
To fill the gap between the wavelength ranges of these two releases, we observed 
our candidates from La Silla (Chile) with 
% the Gamma--Ray Burst Optical Near--infrared Detector 
GROND,  simultaneously in 3 NIR and 4 optical filters.
The partial superposition of the GROND optical filters with the SDSS ones, 
allows to reduce the problems due to the possible variability of the objects.
GROND data are very important, because we expect that the peak of 
the accretion disc emission falls between this wavelength range and the optical 
band provided by the SDSS.
Since only 19 out 
of the 31 radio--loud high--$z$ quasars of the original sample
are visible from La Silla, we reduce our sample to 
these 19, listed in Tab. \ref{19}. 
% with a complete coverage of the 
% optical--NIR wavelength range, that allows a reliable estimate of their black hole masses.
These 19 sources have a very complete IR--optical coverage, and we 
focus on them to estimate the $M_{\rm BH}$ and also to compile a sample of 
good blazar candidates.
% with a mass estimate.

%------------------------------------------------
\begin{table*} 
\centering
\begin{tabular}{lcccccccccccrlllll}
\hline
\hline
RA DEC &$z$ &$R$ &$\log \nu_{\rm r} F_{\nu_{\rm r}}$ % (1.4{\rm GHz})$ 
&$\log F_{\rm X}$ &$\Gamma_{\rm X}$ &Ref %[0.3-10{\rm keV}]
&$\log L_{\rm bol}$ &$\log L_{\rm CIV}$ 
&$\log M_{\rm S11}$ &$\log L_{\rm BLR}$ \\ %
%      & & &[Jy] & & & & & & \\
~[1]      &[2] &[3] &[4] &[5] &[6] &[7] &[8] &[9] &[10] &[11]\\
\hline
  021043.16 --001818.4 &4.733 &140  &--15.87 &--13.07 &2.03 &Sh05  
  					&47.21 &45.20 &-  &46.15\\
  030437.21 +004653.5  &4.305 &2410 &--15.43 &-       &-    &- &46.91 &44.71 &9.65  &45.65\\
  081333.32 +350810.8  &4.922 &610  &--15.78 &-       &-    &- &47.25 &45.76 &10.65 &46.71\\
  083946.22 +511202.8  &4.390 &285  &--15.23 &--12.82 &1.64 &Ba04  
  					&47.53 &45.00 &9.95  &45.94\\
  085111.59 +142337.7  &4.307 &270  &--15.64 &-       &-    &- &47.17 &44.11 &-     &45.05\\
  091316.55 +591921.6  &5.122 &3599 &--15.61 &--14.32 &1.64 &Ba04  
  					&-     &-     &-     &-\\
  091824.38 +063653.3  &4.192 &189  &--15.43 &--12.96 &1.80 &Ba04  
  					&47.21 &44.76 &9.83  &45.71\\
  100645.58 +462717.2  &4.444 &131  &--16.03 &-       &-    &- &47.06 &44.43 &-     &45.36\\
  102623.61 +254259.5  &5.304 &5222 &--14.47 &--12.77 &1.10 &Sb12  
  					&-     &-     &-     &-\\
  103717.72 +182303.0  &4.051 &214  &--15.72 &-       &-    &- &46.75 &45.01 &8.45  &45.95\\
  105320.42 --001649.7 &4.304 &149  &--15.71 &-       &-    &- &47.2  &45.42 &8.83  &46.36\\
  111856.15 +370255.9  &4.025 &155  &--16.05 &-       &-    &- &47.26 &45.13 &9.16  &46.08\\
  114657.79 +403708.6  &5.005 &1013 &--15.76 &-       &-    &- &-     &-     &-     &-\\
  123142.17 +381658.9  &4.137 &264  &--15.47 &-       &-    &- &46.74 &44.91 &8.46  &45.86\\
  123503.03 --000331.7 &4.723 &1493 &--15.59 &-       &-    &- &46.68 &44.60 &9.18  &45.55\\
  124230.58 +542257.3  &4.73  &631  &--15.55 &-       &-    &- &47.10 &44.94 &-     &45.89\\
  130738.83 +150752.0  &4.111 &317  &--15.76 &-       &-    &- &47.09 &44.66 &9.62  &45.61\\
  130940.70 +573309.9  &4.268 &133  &--15.80 &--13.19 &1.80 &Ba04  
  					&47.08 &45.05 &8.32  &45.99\\
  131121.32 +222738.6  &4.612 &394  &--16.04 &-       &-    &- &46.79 &44.89 &9.29  &45.84\\
  132512.49 +112329.7  &4.412 &879  &--15.00 &--13.20 &1.60 &Ba04  
  					&47.44 &45.42 &9.46  &46.37\\
  141209.96 +062406.9  &4.467 &852  &--15.22 &-       &-    &- &47.25 &44.29 &9.86  &45.24\\
  142048.01 +120545.9  &4.034 &1904 &--14.91 &-       &-    &- &47.05 &44.97 &9.28  &45.91\\
  143003.96 +144354.8  &4.834 &391  &--16.51 &-       &-    &- &46.85 &44.79 &10.08 &45.74\\
  143413.05 +162852.7  &4.195 &121  &--16.16 &-       &-    &- &47.08 &45.03 &9.17  &45.97\\
  145212.86 +023526.3  &4.889 &365  &--16.21 &-       &-    &- &46.90 &-     &-     &- \\
  151002.92 +570243.3  &4.309 &13463&--14.45 &--12.27 &1.51 &Yo09  
  					&47.08 &44.87 &8.51 &45.82\\
  152028.14 +183556.1  &4.123 &104  &--16.01 &-       &-    &- &47.07 &44.61 &9.28  &45.43\\
  165913.23 +210115.8  &4.784 &637  &--15.39 &-       &-    &- &47.10 &44.12 &-     &45.06\\
  172026.68 +310431.6  &4.669 &306  &--15.83 &-       &-    &- &46.96 &44.80 &9.73  &45.74\\
  172007.19 +602824.0  &4.424 &124  &--16.15 &-       &-    &- &46.98 &44.84 &8.85  &45.79\\
  222032.50 +002537.5  &4.205 &4521 &--14.87 &-       &-    &- &46.93 &45.05 &9.15  &46.00\\
\hline
\hline
\end{tabular}
\vskip 0.4 true cm
\caption{Sources from the DR7 Quasar Catalog with $z>4$ and radio--loudness $R>100$.
Col. [1]: right ascension and declination (i.e.\ SDSS name: ``SDSS J\ldots");
Col. [2]: redshift;
Col. [3]: radio--loudness defined as in S11 ($R=F_{\rm 5GHz}/F_{\rm 2500\AA}$, fluxes rest frame);
Col. [4]: logarithm of the radio flux in $\ergcms$;
Col. [5]: logarithm of the X--ray flux in the energy range 0.3--10keV, in $\ergcms$;
Col. [6]: X--ray photon index;
Col. [7]: references for X--ray data. 
	     Ba04 stands for Bassett et al.\ (2004),
	     Sb12 for Sbarrato et al.\ (2012b),
	     Sh05 for Shemmer et al.\ (2005), 
	     Yo09 for Young, Elvis \& Risaliti (2009).
Col. [8]: logarithm of the bolometric luminosity, as obtained by S11 ($\ergs$);
Col. [9]: logarithm of the CIV luminosity, as obtained by S11 ($\ergs$);
Col. [10]: logarithm of the BH mass, as derived by S11 (solar masses);
Col. [11]: logarithm of the Broad Line Region luminosity, 
	       calculated from the CIV line luminosity ($\ergs$).
}
\label{31}
\end{table*}
%------------------------------------------------

\section{GROND observations and data analysis}
\label{grond}

The 7--band  GROND imager is mounted at the 2.2m telescope of the
Max-Planck-Gesellschaft (MPG), operated by MPG and ESO
(European Southern Observatory) at La Silla (Chile). 
GROND is able to observe {\it simultaneously} in 7 filters,
from the NIR $K_{\rm s}$ (2400 nm) to the $g^\prime$ band (450 nm), 
i.e.\ from 2200 to 360 nm.

We carried out observations for all sources simultaneously in all 
7 $g^\prime, r^\prime, i^\prime, z^\prime, J, H, K_{\rm s}$ bands. 
The log of the GROND observations and the related observing conditions are reported 
in Tab. \ref{grondlog}. 

The GROND optical and NIR image reduction and photometry were
performed using standard IRAF tasks (Tody 1993), similar to the
procedure described  in Kr\"uhler et al. (2008). 
A general model for the point--spread function (PSF) of each image was
constructed using bright field stars, and it was then fitted to the point source. 
The fields of all sources in our sample are covered  by the SDSS (Smith et al. 2002) survey, 
therefore the absolute calibration of 
the $g^\prime, r^\prime, i^\prime, z^\prime$ bands was  obtained with respect to 
the magnitudes of SDSS stars within the blazar field. 
For all sources the  $J, H, K_{\rm s}$ 
bands calibrations were obtained with respect to magnitudes of 
the Two Micron All Sky Survey (2MASS) stars (Skrutskie et al. 2006).

Tab. \ref{grondlog} reports the log of the GROND observations
while Tab. \ref{grond_tab} reports the observed AB magnitudes, not corrected for the 
Galactic extinction listed in the last column and taken from 
Schlafly \& Finkbeiner (2011).

%------------------------------------------------
\begin{table} 
\centering
\begin{tabular}{lcccccrrllll}
\hline
\hline
RA DEC &$z$ &$\log L_{\rm d}$ &$\log M_{\rm S11}$ &$\log M_{\rm SED}$ 
\\
%      & & &[Jy] & & & & & & \\
~[1]      &[2] &[3] &[4] &[5] \\
\hline
  021043.16 -001818.4 & 4.733 & 46.98 &-     & 9.25$\pm$0.25\\
  030437.21 +004653.5 & 4.305 & 46.53 &9.65  & 9.30$\pm$0.30\\
%  081333.32 +350810.8 & 4.922 & 47.096 & 9.70 & 0.20\\
%  083946.22 +511202.8 & 4.39 & 47.243 & 9.75 & 0.20\\
  085111.59 +142337.7 & 4.307 & 46.89 &-     & 9.45$\pm$0.30\\
%  091316.55 +591921.6 & 5.122 & 9.45 & 0.30 & 46.544\\
  091824.38 +063653.3 & 4.192 & 47.07 &9.83 & 9.75$\pm$0.20\\
%  100645.58 +462717.2 & 4.444 & 9.30 & 0.50 & 46.778\\
  102623.61 +254259.5 & 5.3035 & 46.95 &-     & 9.45$\pm$0.20\\
  103717.72 +182303.0 & 4.051 & 46.56 &8.45 & 9.48$\pm$0.20\\
  105320.42 -001649.7 & 4.304 & 47.04 &8.83 & 9.00$\pm$0.30\\
%  111856.15 +370255.9 & 4.025 & 9.85 & 0.20 & 46.978\\
%  114657.79 +403708.6 & 5.005 & 9.65 & 0.20 & 47.021\\
%  123142.17 +381658.9 & 4.137 & 9.40 & 0.25 & 46.623\\
  123503.03 -000331.7 & 4.723 & 46.52 &9.18 & 9.50$\pm$0.30\\
%  124230.58 +542257.3 & 4.73 & 9.55 & 0.20 & 46.813\\
  130738.83 +150752.0 & 4.111 & 46.78 &9.62 & 9.25$\pm$0.20\\
%  130940.70 +573309.9 & 4.268 & 9.95$\pm$0.20 & 47.061\\
  131121.32 +222738.6 & 4.612 & 46.60 &9.29 & 9.55$\pm$0.30\\
  132512.49 +112329.7 & 4.412 & 47.11 &9.46 & 9.50$\pm$0.30\\
  141209.96 +062406.9 & 4.467 & 47.02 &9.86 & 8.95$\pm$0.20\\
  142048.01 +120545.9 & 4.034 & 46.73 &9.28 & 9.30$\pm$0.25\\
  143003.96 +144354.8 & 4.834 & 46.65 &10.08 & 9.30$\pm$0.25\\
  143413.05 +162852.7 & 4.195 & 46.78 &9.17 & 9.00$\pm$0.30\\
  145212.86 +023526.3 & 4.889 & 46.67 &-        & 9.40$\pm$0.30\\
%  151002.92 +570243.3 & 4.309 & 9.48 & 0.30 & 46.771\\
  152028.14 +183556.1 & 4.123 & 46.81 &9.28 & 9.10$\pm$0.30\\
  165913.23 +210115.8 & 4.784 & 46.65 &-        & 9.60$\pm$0.30\\
%  172026.68 +310431.6 & 4.669 & 9.95 & 0.20 & 46.929\\
%  172007.19 +602824.0 & 4.424 & 9.20 & 0.20 & 46.700\\
  222032.50 +002537.5 & 4.205 & 46.65 &9.15 & 9.30$\pm$0.25\\
\hline
\hline
\end{tabular}
\vskip 0.4 true cm
\caption{Sources from the DR7 Quasar Catalog with $z>4$ and radio--loudness $R>100$, 
seen by GROND.
Col. [1]: right ascension and declination (i.e.\ SDSS name: ``SDSS J\ldots");
Col. [2]: redshift;
Col. [3]: logarithm of the disc luminosity in $\ergs$, as derived in this work;
Col. [4]: logarithm of the BH mass, as derived by S11 (solar masses);
Col. [5]: logarithm of the BH mass as derived in this work (solar masses), 
	      with the confidence interval.
}
\label{19}
\end{table}
%------------------------------------------------

% ------------------------------------------------------------------------
\begin{table*} 
\centering
\begin{tabular}{llllll }
\hline
\hline
Name       &Date         &Mid time      &Exp: opt/IR &Seeing  &Mean airmass \\
           &yyyy--mm--dd &$[{\rm UTC}]$  &[s]      &[arcsec]        &                \\
\hline   
SDSS J021043.16-001818.4 & 2012-08-10 & 10:10:58  & 919/960 & 1.3 & 1.16\\ 
SDSS J030437.21+004653.6 & 2012-08-21 & 09:12:29 & 602/720 & 0.9 & 1.17\\
SDSS J085111.60+142337.8 & 2012-04-20 & 23:27:28 & 919/960 & 1.3 & 1.38\\
SDSS J091824.38+063653.3 & 2012-04-21 & 23:27:00 & 919/960 & 1.4 & 1.25\\
SDSS J102623.62+254259.6 & 2012-04-16 & 23:29:13 & 919/960 & 1.2 & 2.11\\
SDSS J103717.73+182303.1 & 2012-06-04 & 23:18:23 & 919/960 & 1.4 & 1.54\\
SDSS J105320.42-001649.5 & 2012-06-08 & 23:22:55 & 919/960 & 1.0 & 1.18\\
SDSS J123503.02-000331.6 & 2012-06-14 & 23:10:40 & 805/840 & 1.3 & 1.16\\
SDSS J130738.83+150752.1 & 2012-07-04 & 23:17:37 & 919/960 & 2.6 & 1.41\\
SDSS J131121.32+222738.6 & 2012-06-30  & 01:07:42 &  919/960 & 1.1 & 1.88\\
SDSS J132512.49+112329.7 & 2012-07-06 & 23:06:23 & 920/960 & 1.1 & 1.32\\
SDSS J141209.96+062406.8 & 2012-08-26 & 23:27:05 & 460/480 & 1.3 & 1.75\\
SDSS J142048.01+120546.0 & 2012-08-12 & 23:34:17 & 919/960 & 2.9 & 1.62\\
SDSS J143003.96+144354.8 & 2012-08-13 & 00:13:27 & 919/960 & 2.0 & 1.82\\
SDSS J143413.05+162852.7 & 2012-08-21 & 23:31:37 & 672/840 & 1.8 & 1.90\\
SDSS J145212.85+023526.4 & 2012-08-28 & 23:33:18 & 919/960 & 1.5 & 1.23\\
SDSS J152028.14+183556.1 & 2012-09-04 & 23:38:49 & 919/960 & 1.8 & 2.04\\
SDSS J165913.23+210115.8 & 2012-09-15 & 23:57:11 & 919/960 & 1.3 & 1.83\\
SDSS J222032.50+002537.5 & 2012-05-28 & 19:35:30 & 460/480 &1.7&1.15\\

\hline
\hline 
\end{tabular}
\vskip 0.4 true cm
\caption{
Log of the GROND observations. 
Exposures refer to optical/NIR filters while the average seeing is calculated in 
the $r^\prime$ band.
} 
\label{grondlog}
\end{table*}
% ------------------------------------------------------------------------

% ------------------------------------------------------------------------
\begin{table*} 
\centering
\begin{tabular}{lllll llll }
\hline
\hline
Name    &$g'$ &$r'$ &$i'$ &$z'$ &$J$ &$H$ &$K_s$  &$A_V$\\
\hline   
SDSS J021043.16-001818.4 & 22.75$\pm$0.13 & 20.30$\pm$0.02 & 19.34$\pm$0.02 & 19.17$\pm$0.02 & 19.28$\pm$0.05 & 19.13$\pm$0.07 & 18.96$\pm$0.14 & 0.080 \\
SDSS J030437.21+004653.6 & 22.80$\pm$0.10 & 20.78$\pm$0.03 & 20.34$\pm$0.04 & 20.32$\pm$0.04 & 20.29$\pm$0.15 & 19.74$\pm$0.18 & 20.29$\pm$0.45 & 0.277 \\
SDSS J085111.60+142337.8 & 22.28$\pm$0.07 & 20.45$\pm$0.02 & 19.86$\pm$0.02 & 19.52$\pm$0.02 & 19.12$\pm$0.07 & 18.93$\pm$0.10 & 18.48$\pm$0.16 & 0.105 \\
SDSS J091824.38+063653.3 & 21.43$\pm$0.04 & 19.63$\pm$0.01 & 19.03$\pm$0.02 & 18.92$\pm$0.02 & 18.53$\pm$0.04 & 18.26$\pm$0.05 & 18.15$\pm$0.16 & 0.119  \\
SDSS J102623.62+254259.6 & 24.51$\pm$0.56 & 22.07$\pm$0.08 & 19.97$\pm$0.03 & 19.86$\pm$0.04 & 19.50$\pm$0.10 & 19.23$\pm$0.15 & 19.07$\pm$0.21 & 0.064 \\ 
SDSS J103717.73+182303.1 & 21.47$\pm$0.01 & 19.93$\pm$0.02 & 19.75$\pm$0.03 & 20.04$\pm$0.03 & 19.63$\pm$0.07 & 19.37$\pm$0.14 & 19.70$\pm$0.28 & 0.073 \\
SDSS J105320.42-001649.5 & 21.93$\pm$0.03 & 19.42$\pm$0.01 & 19.35$\pm$0.01 & 19.21$\pm$0.02 & 18.92$\pm$0.05 & 18.58$\pm$0.06 & 18.22$\pm$0.09 & 0.146 \\
SDSS J123503.02-000331.6 & 24.47$\pm$0.31 & 21.25$\pm$0.03 & 21.48$\pm$0.03 & 20.32$\pm$0.04 & 20.11$\pm$0.15 & 19.69$\pm$0.29 & 20.38$\pm$0.69 & 0.062 \\
SDSS J130738.83+150752.1 & 21.20$\pm$0.06 & 19.99$\pm$0.02 & 19.84$\pm$0.04 & 19.77$\pm$0.04 & 19.43$\pm$0.10 & 19.32$\pm$0.12 & 19.25$\pm$0.16 & 0.103 \\
SDSS J131121.32+222738.6 &    $>23.5$           & 21.29$\pm$0.04 & 20.60$\pm$0.05 & 20.43$\pm$0.05 & 20.08$\pm$0.13 & 19.51$\pm$0.14 & 19.54$\pm$0.28 & 0.035 \\
SDSS J132512.49+112329.7 & 22.15$\pm$0.06 & 19.51$\pm$0.01 & 19.35$\pm$0.01 & 18.94$\pm$0.02 & 19.02$\pm$0.05 & 18.56$\pm$0.06 & 18.63$\pm$0.09 & 0.065 \\
SDSS J141209.96+062406.8 & 22.11$\pm$0.19 & 20.18$\pm$0.03 & 19.54$\pm$0.03 & 19.49$\pm$0.04 & 19.56$\pm$0.09 & 19.50$\pm$0.13 &  $>19.4$ & 0.089 \\
SDSS J142048.01+120546.0 & 21.21$\pm$0.03 & 19.93$\pm$0.02 & 19.55$\pm$0.03 & 19.69$\pm$0.03 & 19.64$\pm$0.11 & 19.39$\pm$0.17 & 19.24$\pm$0.40 & 0.076 \\
SDSS J143003.96+144354.8 & $>24.2$              & 21.67$\pm$0.06 & 20.31$\pm$0.03 & 20.11$\pm$0.04 & 19.56$\pm$0.09 & 19.58$\pm$0.15 & 19.26$\pm$0.18 & 0.072 \\
SDSS J143413.05+162852.7 & 22.09$\pm$0.15 & 20.14$\pm$0.02 & 20.13$\pm$0.05 & 20.23$\pm$0.07 & 19.74$\pm$0.12 & 19.71$\pm$0.20 & 19.75$\pm$0.31 & 0.066 \\
SDSS J145212.85+023526.4 & $>22.9$              & 21.96$\pm$0.11 & 20.11$\pm$0.04 & 20.00$\pm$0.05 & 19.82$\pm$0.12 & 19.81$\pm$0.18 & 20.22$\pm$0.92 & 0.113 \\
SDSS J152028.14+183556.1 & 21.21$\pm$0.05 & 19.74$\pm$0.02 & 19.41$\pm$0.02 & 19.32$\pm$0.03 & 19.41$\pm$0.11 & 19.33$\pm$0.15 & 19.23$\pm$0.29 & 0.148 \\  
SDSS J165913.23+210115.8 & $>24.3$              & 21.57$\pm$0.04 & 20.10$\pm$0.02 & 19.99$\pm$0.03 & 19.71$\pm$0.09 & 18.85$\pm$0.08 & 19.63$\pm$0.33 & 0.202 \\
SDSS J222032.50+002537.5 & 22.31$\pm$0.10 & 20.24$\pm$0.02 & 19.67$\pm$0.06 & 20.01$\pm$0.04 & 19.67$\pm$0.12 & 19.56$\pm$0.19 & 19.32$\pm$0.58 & 0.183 \\
\hline
\hline 
\end{tabular}
\vskip 0.4 true cm
\caption{Observed magnitudes, not corrected for Galactic foreground reddening, in the AB system.
Errors include systematics.
The last column reports the value of the Galactic $A_V$ from Schlafly \& Finkbeiner (2011).
}
\label{grond_tab}
\end{table*}
% ------------------------------------------------------------------------

\section{The IR to optical SED}
\label{sed}

In the Appendix we show the IR to optical SED of all the 18 sources in our
sample.
We did not include SDSS J102623.61+254259.5, because we already applied 
to this source the accretion disc fitting method in our previous work 
(Sbarrato et al.\ 2012b).
The first evident feature of all the SEDs is the Ly$\alpha$ forest, i.e.\ 
the prominent  absorption clearly visible in 
the SDSS spectra (black line) 
and in the $g^\prime$ and sometimes $r^\prime$ band GROND photometric data 
%and in the last one or two GROND points 
(red points) at frequencies larger than $\log(\nu/{\rm Hz})=15.4$ 
(i.e.\ after the Ly$\alpha$ line).
The Ly$\alpha$ forest is caused by  intervening Ly$\alpha$ 
clouds, present in random density and distance along each line of sight.
Since the effects of this absorption are unlikely to be predicted, 
we do not try to fit the data at frequencies larger than the Ly$\alpha$.
[blue and cyan lines in Fig. \ref{sed1}--\ref{sed6} become dashed at $\nu$ larger than
$\log(\nu/{\rm Hz})=15.4$].

Another recurrent feature is the increase in luminosity 
in the redder WISE bands 
%of the lowest frequency data obtained from WISE 
with respect to the decreasing tendency that the other IR data have 
toward lower frequencies.
Where the lowest frequency WISE data is not an upper limit, indeed, 
it has always a luminosity at least half a decade larger than one would expect 
extrapolating the 
flux from larger frequencies.
This could be ascribed to the presence of a different component, such as 
the synchrotron hump or an emission from the dusty torus (Calderone, Sbarrato \& Ghisellini 2012).

The most useful feature that characterizes the optical--NIR spectra of our sources 
is the possibility to directly observe the $\nu L_\nu$ peak of the accretion disc spectrum.
% As can be seen from the SEDs, 
Just before the Ly$\alpha$ forest 
absorption the SDSS data highlight a peak in the $\nu L_\nu$ emitted spectrum, 
that ends the clear increasing trend of the flux from GROND.
The highest redshift objects show this peak even in the GROND bands.
%frequencies observed by GROND itself.
This was already pointed out in Sbarrato et al.\ 2012b, 
where GROND observations allowed to clearly draw 
%where the GROND data were obtained specifically to draw more clearly 
the peak of the thermal emission 
of B2 1023+25 ($z=5.3$).

The visibility of the accretion disc, along with its $\nu L_\nu$ peak, 
allow us to fit it with a theoretical 
model, in order to derive some  basic properties, such as the overall disc luminosity 
$L_{\rm d}$ and the black hole mass $M_{\rm BH}$.

\subsection{The disc Luminosity}

One way to derive the overall disc luminosity is through the total 
luminosity of the Broad Line Region (BLR).
The accretion disc is the principal source of UV photons that ionize the BLR, 
hence the two luminosities have to be strictly related.
The BLR re--emits a fraction $C$ of the ionizing luminosity, that corresponds to 
the covering factor of the BLR with respect to the disc.
The average value of $C$ is 0.1 (Baldwin \& Netzer, 1978; Smith et al.\ 1981), with 
a large dispersion depending on the geometrical features of the BLR itself.
The luminosity emitted from the whole BLR can be derived from the luminosities of the 
principal broad emission lines through the templates calculated by Francis et al.\ (1991) 
and Vanden Berk et al.\ (2001).
In this way we obtain a range of possible values of the disc luminosity,  
determined by the uncertainties on the covering factor and on the templates.

On the other hand, when the peak of the accretion disc component is
directly visible, we can strongly reduce the uncertainties on its bolometric
luminosity.
This often occurs for the sources in our sample, despite the possible contribution of
the jet synchrotron emission.
This is due to the large redshift of the source, red--shifting the peak of the
disc component within the optical band, red--ward of the Lyman--$\alpha$ limit,
and because our sources, being very powerful, all have a synchrotron component
peaking in the far IR or mm band.
Hence, with the good coverage of the optical--NIR band given by the SDSS, 
GROND and WISE data, we obtain for all our blazar candidates the peak luminosities 
and can constrain the total disc luminosity. 
The uncertainty of our method is hence associated only 
to the measurement errors of the flux and to its possible variability, since the
GROND, WISE and SDSS data are not simultaneous.

The upper panel of Fig.\ \ref{MLd} shows the direct comparison of our estimates 
of the disc luminosity with half the bolometric luminosity as defined in Richards et al.\ (2006).
Note that no peculiar trend nor an evident offset is present between the two luminosities, 
but they are clearly linearly correlated. 
This represents a further confirmation of the ``naked" accretion disc.
Indeed, the bolometric luminosity in Richards et al.\ (2006) is derived from the observed flux 
measured at specific observed wavelengths.
If the flux was a combination of the emissions from 
the accretion disc and an additional component, 
e.g.\ the synchrotron emission, it would not be a good tracer of the accretion disc emission. 
In that case, it would not be so tightly correlated with our estimate of the overall 
accretion disc luminosity, that suffers less from possible contaminations, being 
directly derived from the peak of the disc emission.

\subsection{Black Hole Mass Estimate}
\label{mass}

The black hole mass is a key feature to describe the quasars in our sample.
The high redshift, however, introduces some difficulties in applying the commonly used 
virial method.
Indeed, the CIV line is present only for $z<4.7$ in the SDSS spectral range, hence it is 
not possible to obtain the virial masses for the whole sample.
Moreover, the CIV line is the less reliable broad emission line to apply 
the virial argument (Shen 2013 and references therein).
Therefore, to derive the $M_{\rm BH}$ coherently for all our objects, we apply a 
different method.

As mentioned above, 
the extreme power of these objects ensure that the 
synchrotron hump is at frequencies low enough to leave ``naked" the spectrum 
produced by the accretion disc, 
that peaks in the optical--IR band due to the very high redshift.
Hence, with a good coverage of this wavelength range, the accretion disc emission 
can be directly modeled and fitted.
Since bright broad Ly$\alpha$ and CIV lines are visible in the optical spectra of our sample, we 
can reasonably assume that the accretion disc is radiatively efficient, being able to ionize 
the broad line region (BLR).
We thus apply the simplest hypothesis for a radiatively efficient accretion disc, 
i.e. the Shakura--Sunyaev disc (Shakura \& Sunyaev, 1973).
According to this model, the geometrically thin, optically thick accretion disc 
can be fitted with a multicolor black body spectrum.
This spectrum depends only on the central black hole mass ($M_{\rm BH}$) and on the 
mass accretion rate ($\dot M$), that is directly traced 
by the total luminosity emitted by the disc ($L_{\rm d}=\eta\dot Mc^2$, with 
$\eta=0.08$\footnote{
The efficiency can be defined as $\eta=GM_{\rm BH}/(2R_{\rm in}c^2)$, where 
$R_{\rm in}$ is the inner radius of the accretion disc, 
for which we assume a value of $R_{\rm in}=6GM_{\rm BH}/c^2$, 
as in the case of a non--rotating black hole. 
With this simple hypothesis, we assume $\eta=0.08$.
}).
In our case, we can precisely constrain both parameters.

The same model has been applied by Calderone et al.\ (2013) to fit the accretion 
discs of a sample of radio--loud Narrow Line Seyfert 1s,
and in that paper the model is described in detail.
% In the Appendix, the authors describe in details the model they applied.
% They considered a canonical Shakura--Sunyaev disc, and find 
% some peculiar features of its spectrum represented in $\nu L_\nu$.
% Specifically, they find a tight relation between 
% the overall disc luminosity and the value of the peak of the emitted spectrum 
% represented in $\nu L_\nu$.
% The value of the peak in this representation (i.e.\ $\nu_{\rm p} L_{\nu_{\rm p}}$) 
% is exactly half of the luminosity emitted by the disc ($L_{\rm d}$).
As we already pointed out, at these redshifts, we clearly see the $\nu L_\nu$ peak
of the accretion disc component. 
This allows to directly derive the disc luminosity $L_{\rm d}$ of all our sources 
[i.e. $L_{\rm d}=2\nu_{\rm p}L_{\nu_{\rm p}}$, where $\nu_{\rm p}$ is the peak
frequency of the disc spectrum].

With the total disc luminosity fixed by the $\nu L_\nu$ peak, we are left with only the 
black hole mass as a free model parameter.
In a multicolor blackbody model, for a fixed value of the disc luminosity, 
the black hole mass affects only the peak frequency, 
since it determines the emitting surface and its temperature.
In other words, a larger black hole mass implies a larger disc surface, that needs to be 
colder to emit a fixed $L_{\rm d}$.
This implies a rigid shift of the $\nu L_\nu$ peak towards smaller frequencies 
(see also Fig.\ 3 in Calderone et al.\ 2013).
Therefore, if the overall disc luminosity is determined through the 
$\nu_{\rm p} L_{\nu_{\rm p}}$, 
it is straightforward to estimate the central
black hole mass, with a good sampling of the optical--NIR energy range.

\begin{figure}
\vskip -0.6 cm 
\hskip -0.4 cm
\psfig{figure=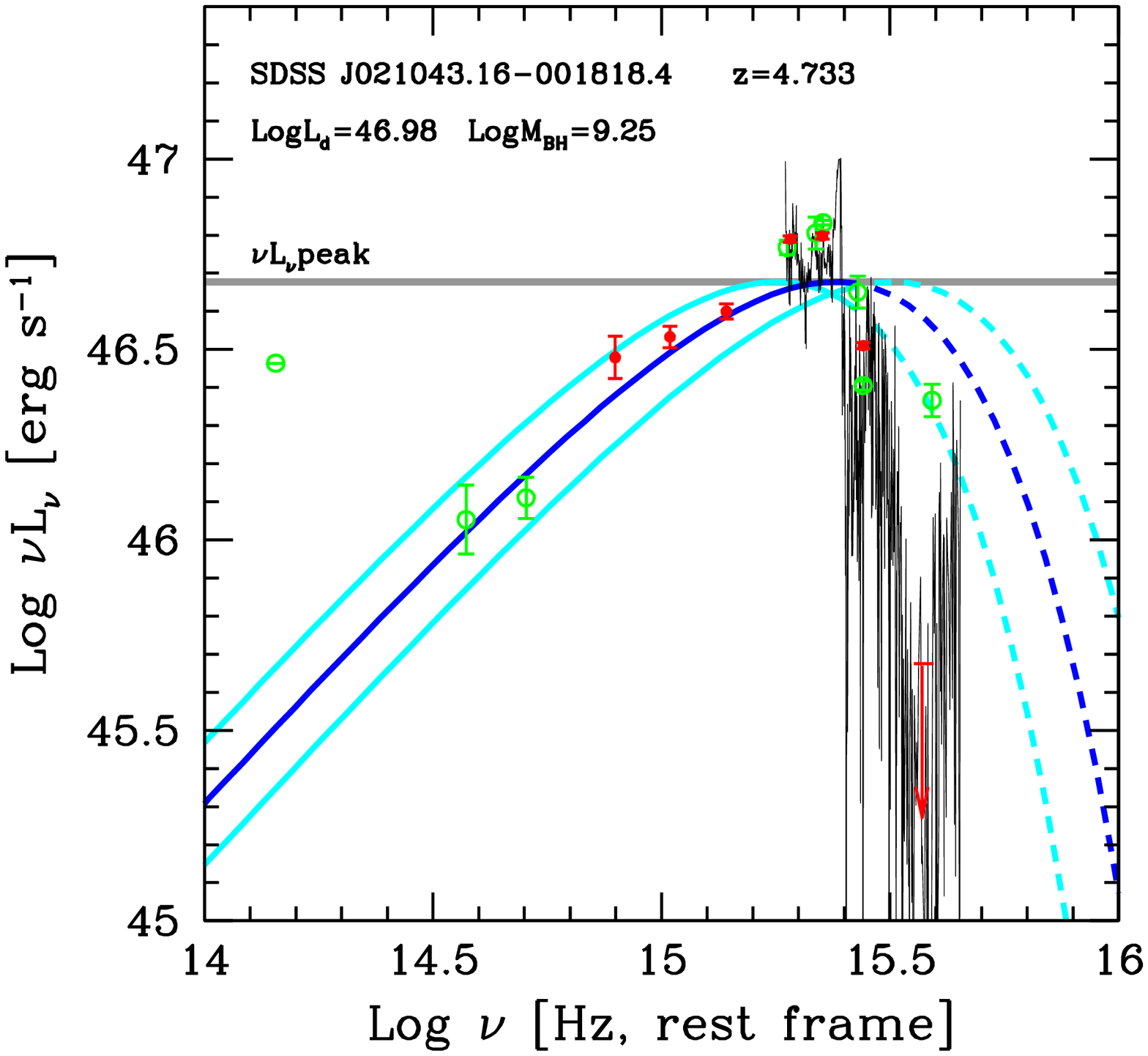,width=9cm,height=8cm}
\vskip -1.1 cm 
\hskip -0.4 cm
\psfig{figure=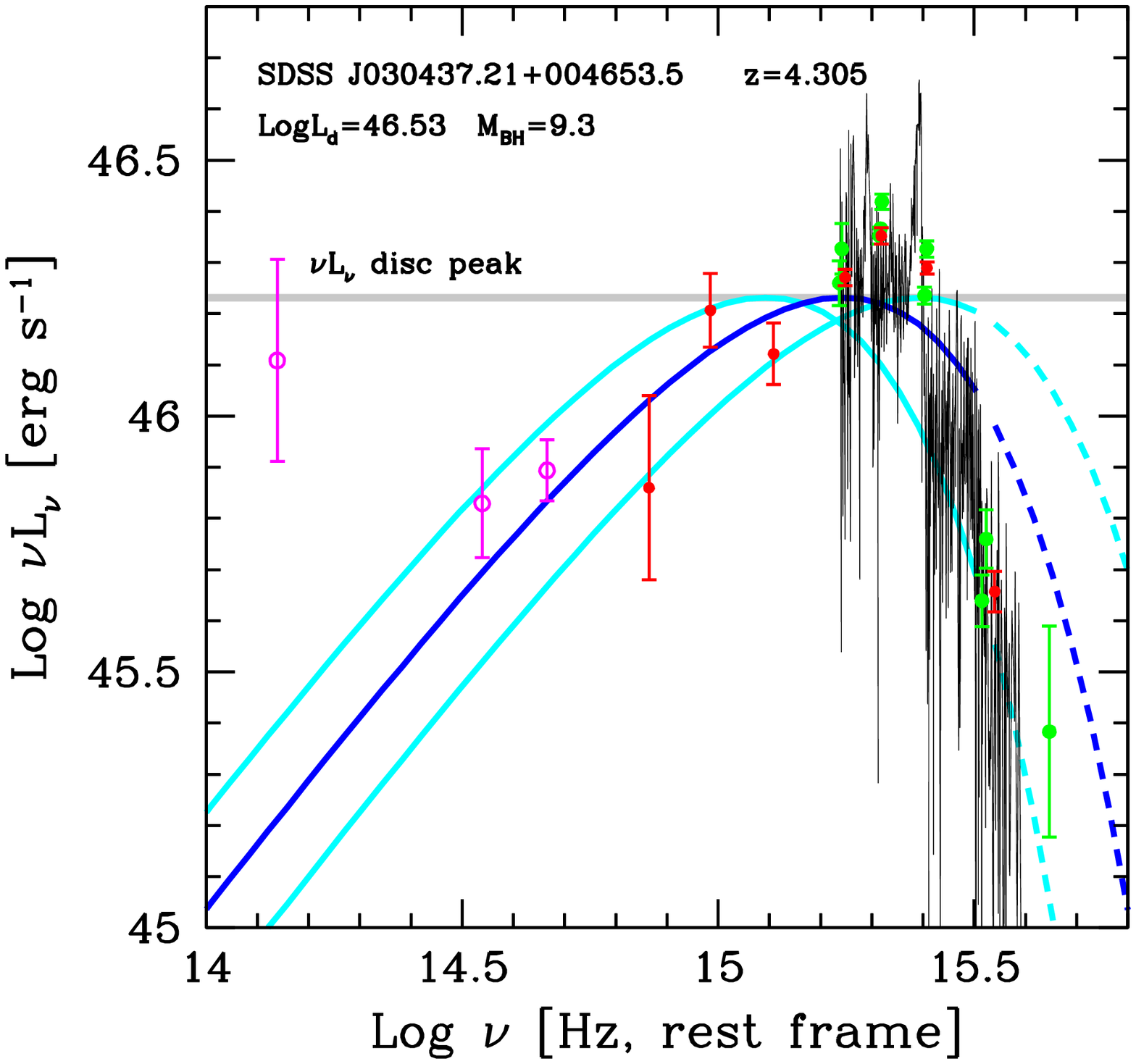,width=9cm,height=8cm}
\vskip -1.1 cm 
\hskip -0.4 cm
\psfig{figure=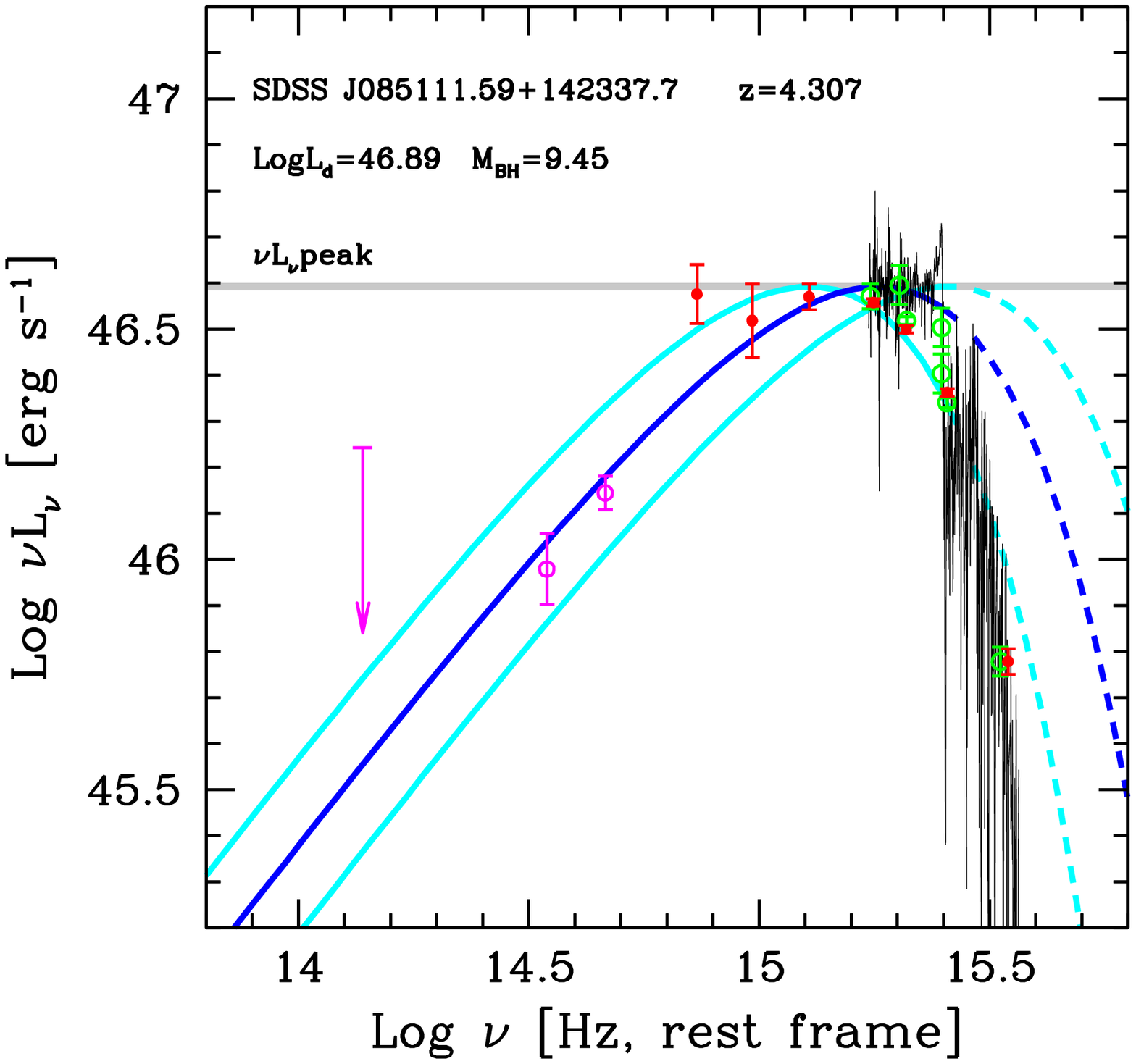,width=9cm,height=8cm}
\vskip -0.5 cm
\caption{
IR--optical--UV SEDs of the quasars observed by GROND in the rest frame, 
together with standard accretion models.
In each figure, the black line is the SDSS spectrum.
Red and magenta points are data from GROND and WISE, respectively.
Green points are archival data (ASDC SED builder, NED).
The grey stripe indicates the $\nu L_\nu$ peak luminosity of the disc.
The blue line is the multicolor black body that best fits our data, calculated with 
$L_{\rm d}$ and $M_{\rm BH}$ values reported in each figure.
Outside the range marked by the cyan lines, the disc model cannot fit 
satisfactory the data.
Note that at $\log\nu>15.4$ the models do not fit the data, because of the 
Ly$\alpha$ forest.
}
\label{sed1}
\end{figure}

\begin{figure}
\vskip -0.6 cm 
\hskip -0.4 cm
\psfig{figure=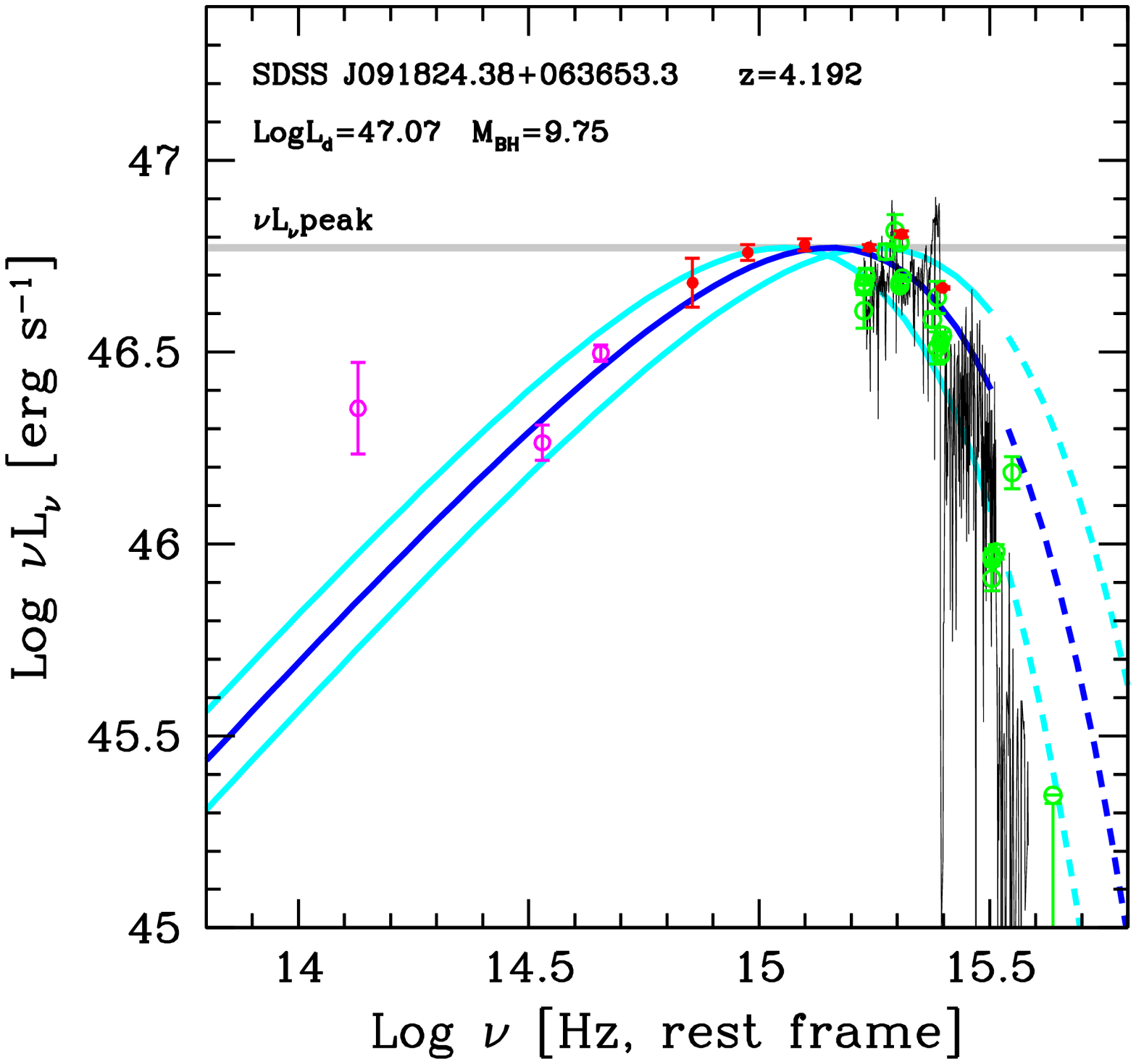,width=9cm,height=8cm}
\vskip -1.1 cm 
\hskip -0.4 cm
\psfig{figure=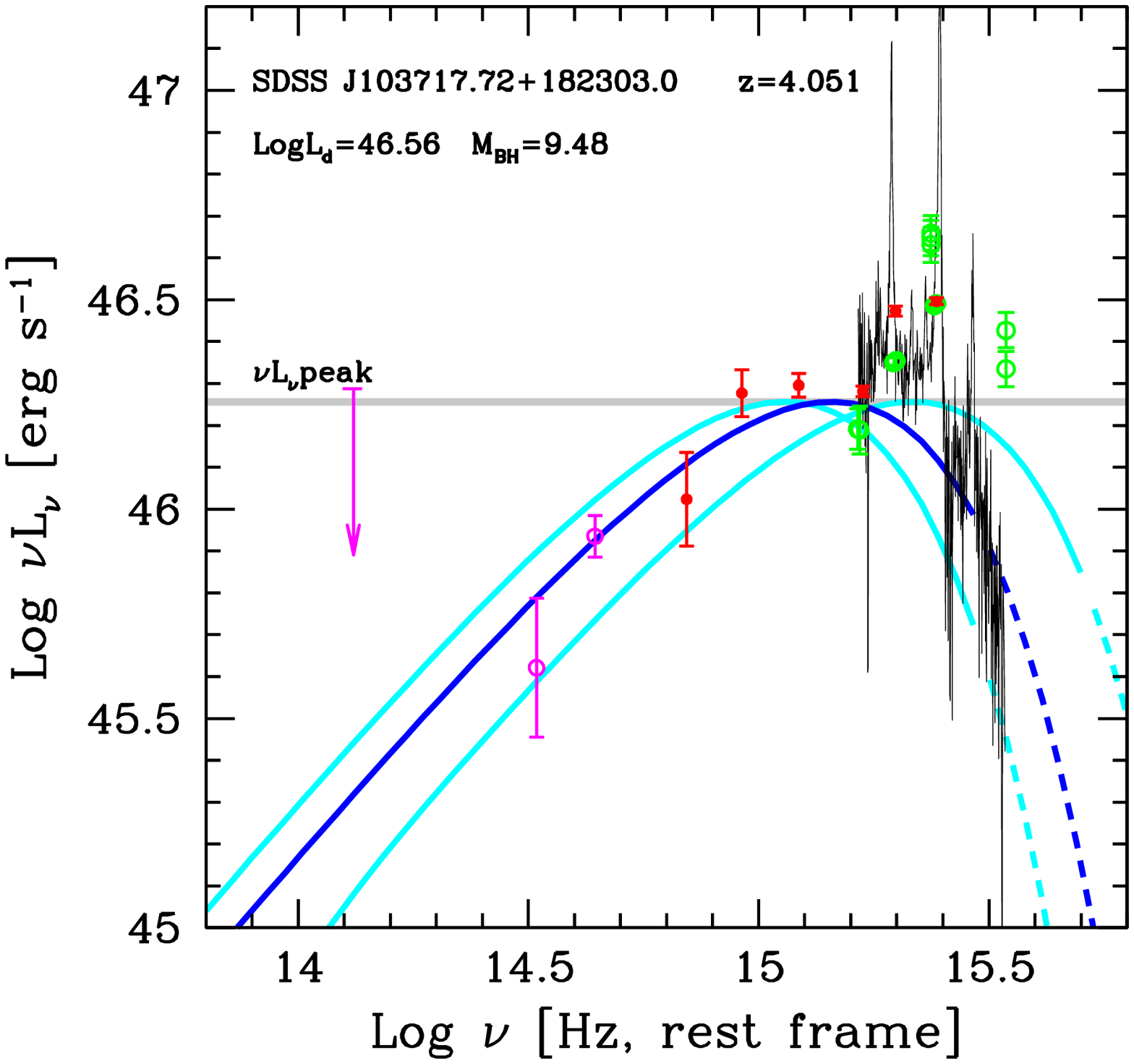,width=9cm,height=8cm}
\vskip -1.1 cm 
\hskip -0.4 cm
\psfig{figure=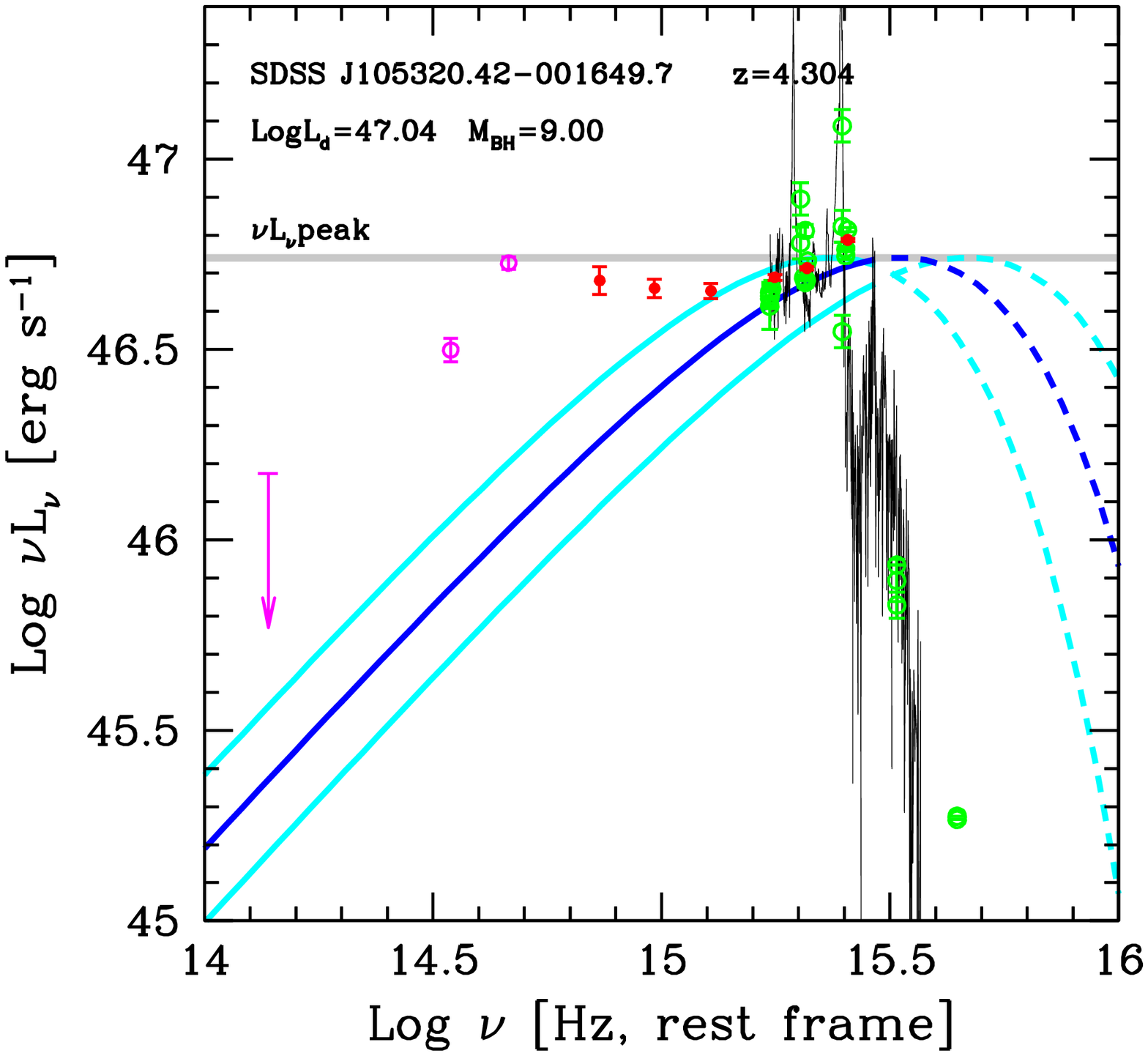,width=9cm,height=8cm}
\vskip -0.5 cm
\caption{
See caption of Fig.\ \ref{sed1}
}
\label{sed2}
\end{figure}

\begin{figure}
\vskip -0.6 cm 
\hskip -0.4 cm
\psfig{figure=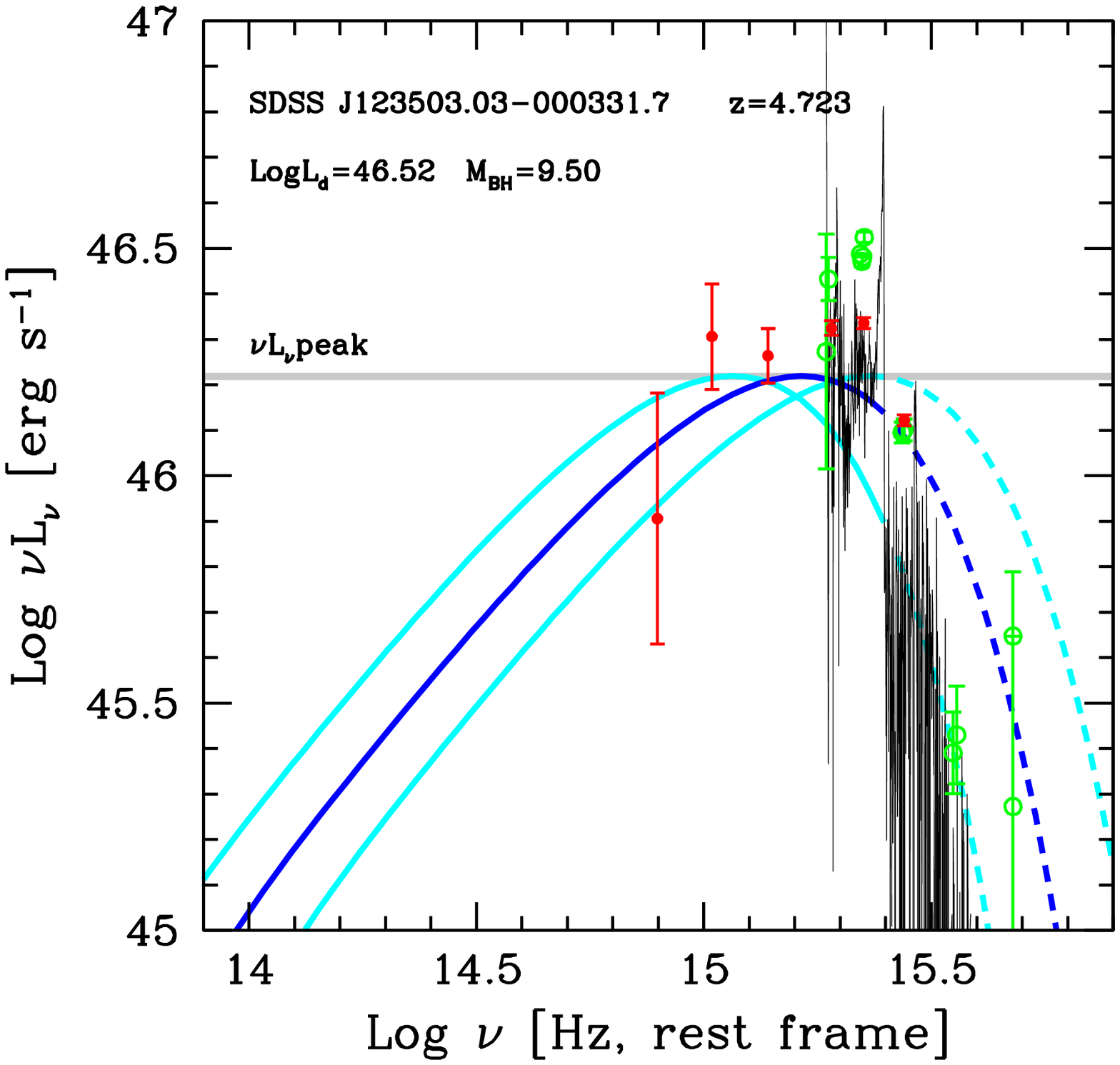,width=9cm,height=8cm}
\vskip -1.1 cm 
\hskip -0.4 cm
\psfig{figure=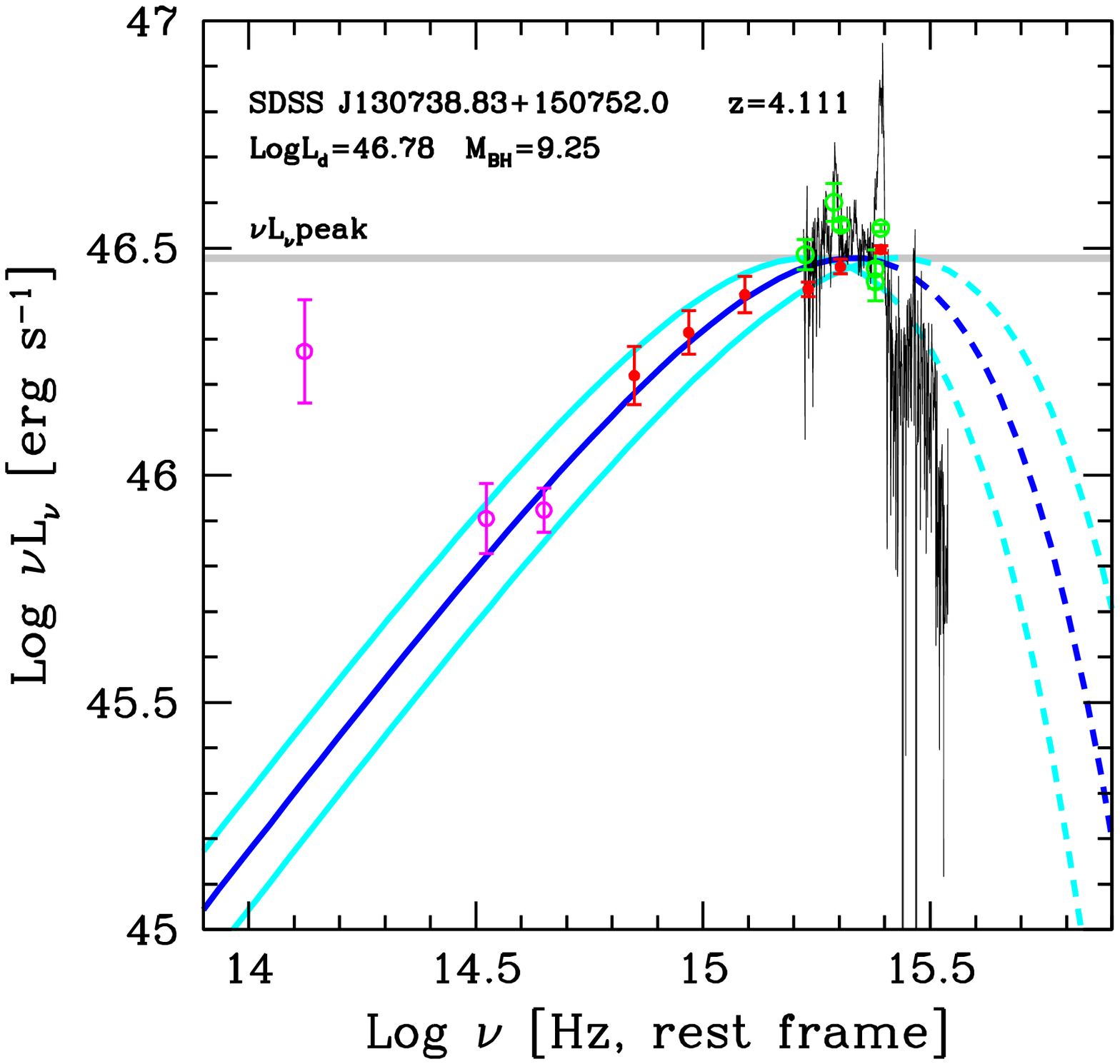,width=9cm,height=8cm}
\vskip -1.1 cm 
\hskip -0.4 cm
\psfig{figure=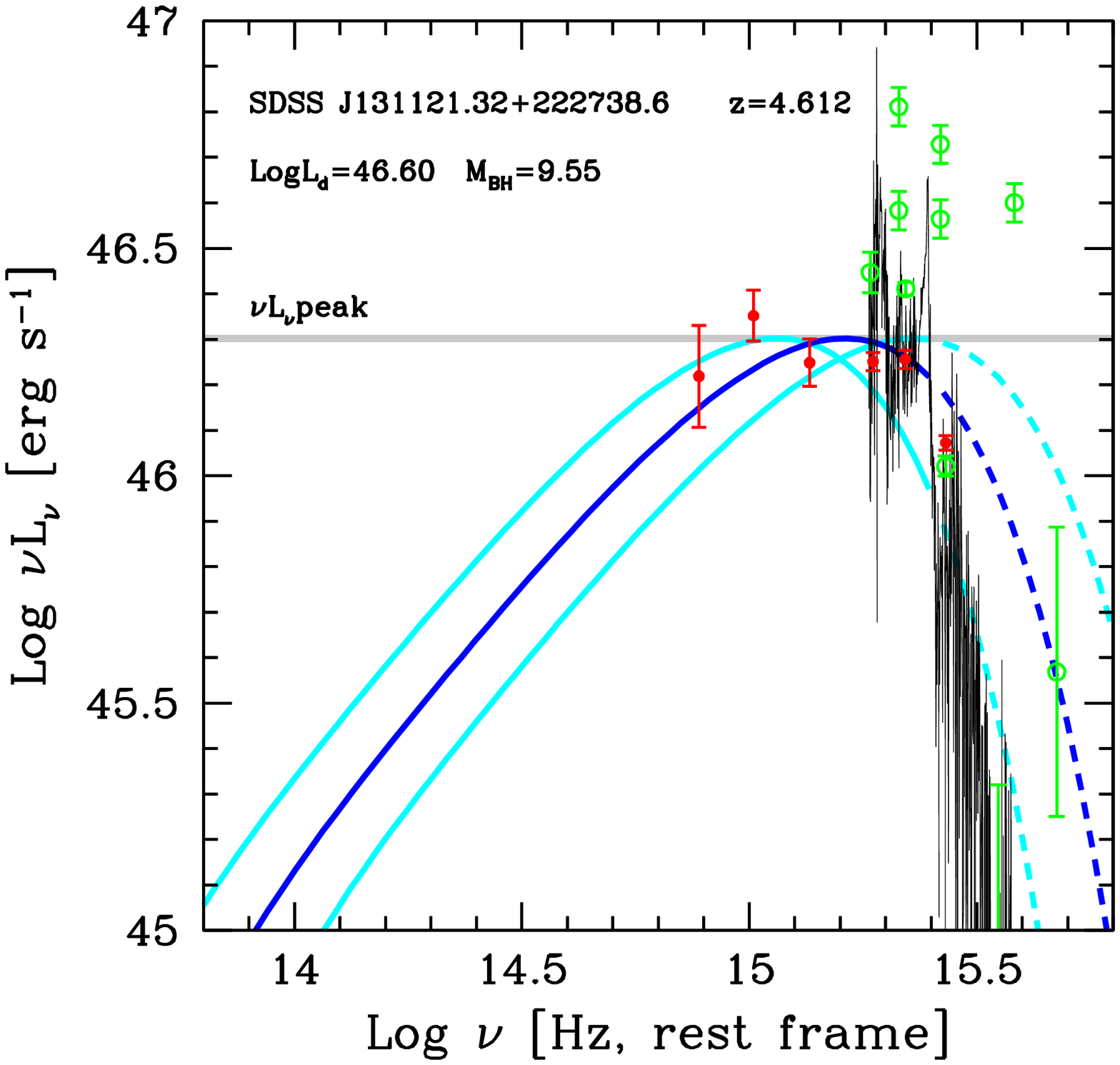,width=9cm,height=8cm}
\vskip -0.5 cm
\caption{
See caption of Fig.\ \ref{sed1}
}
\label{sed3}
\end{figure}

\begin{figure}
\vskip -0.6 cm 
\hskip -0.4 cm
\psfig{figure=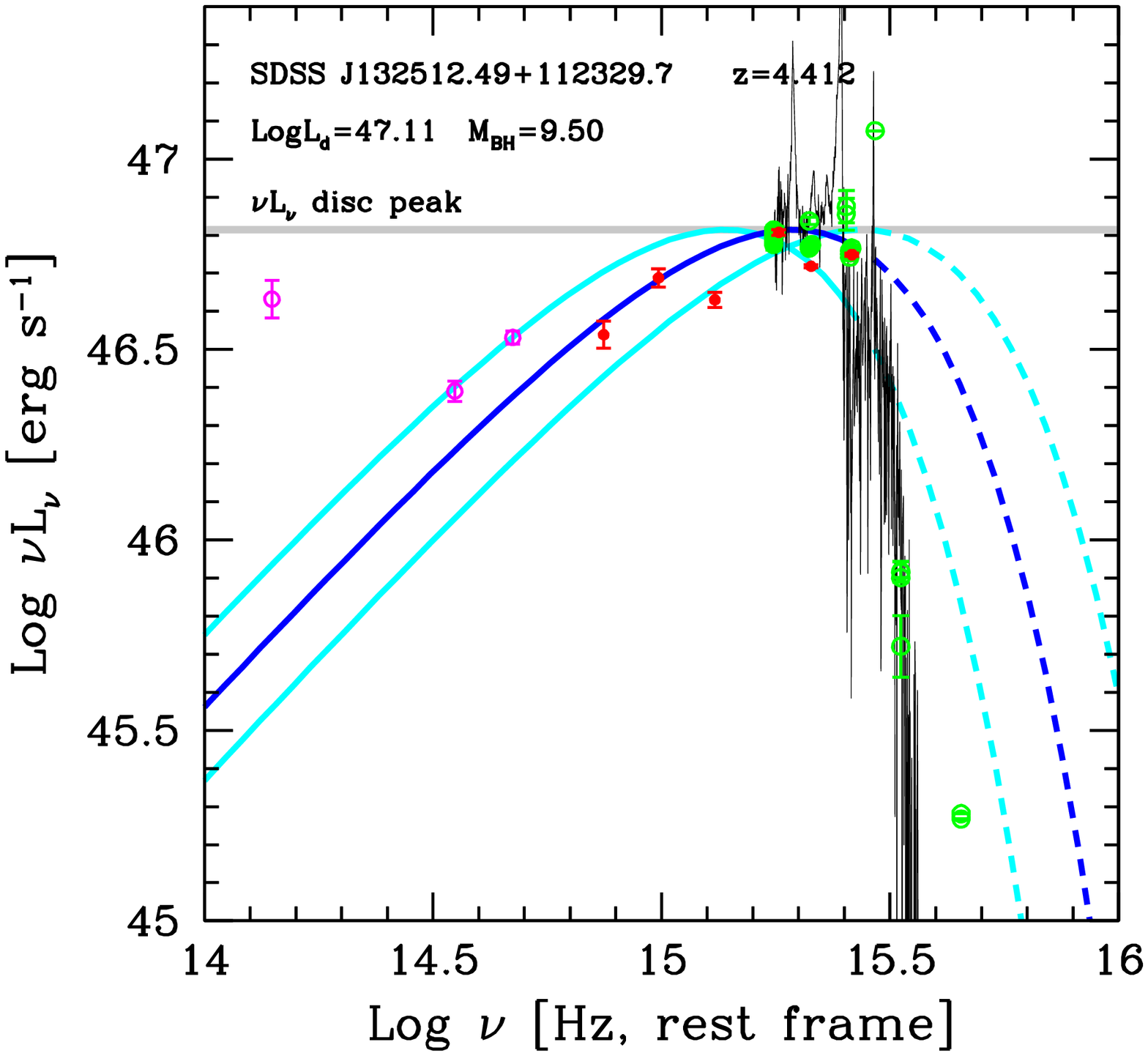,width=9cm,height=8cm}
\vskip -1.1 cm 
\hskip -0.4 cm
\psfig{figure=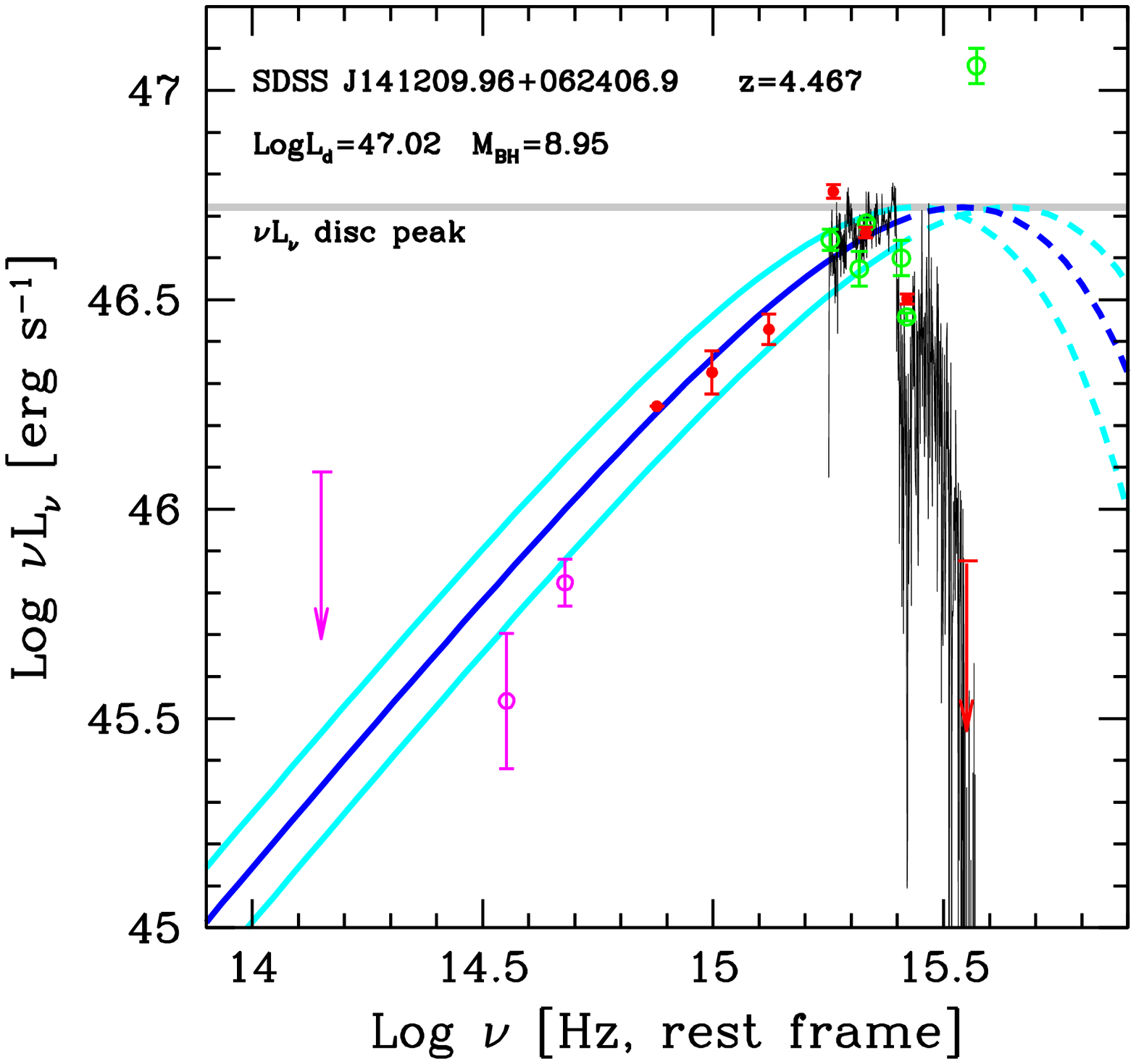,width=9cm,height=8cm}
\vskip -1.1 cm 
\hskip -0.4 cm
\psfig{figure=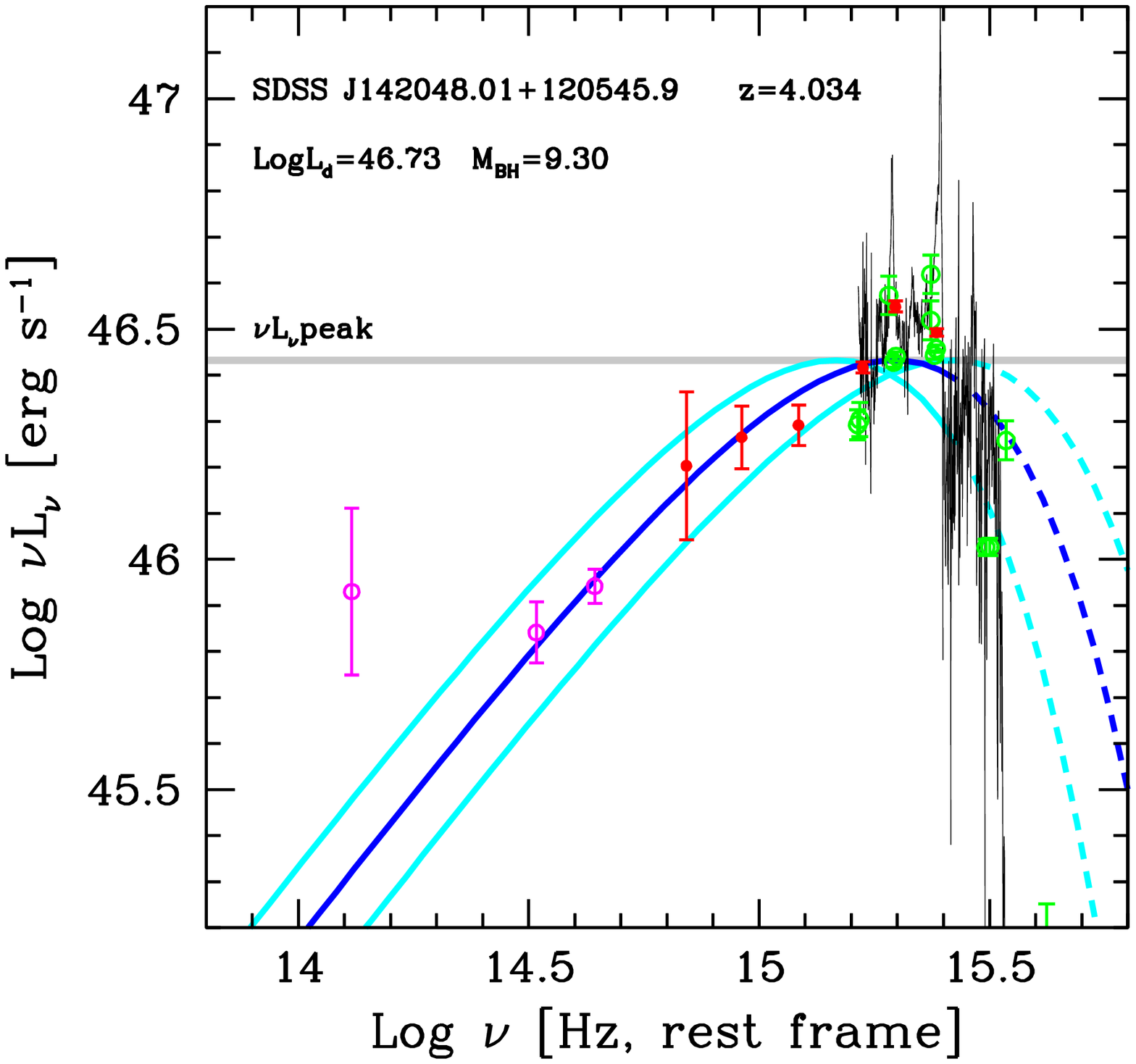,width=9cm,height=8cm}
\vskip -0.5 cm
\caption{
See caption of Fig.\ \ref{sed1}
}
\label{sed4}
\end{figure}

\begin{figure}
\vskip -0.6 cm 
\hskip -0.4 cm
\psfig{figure=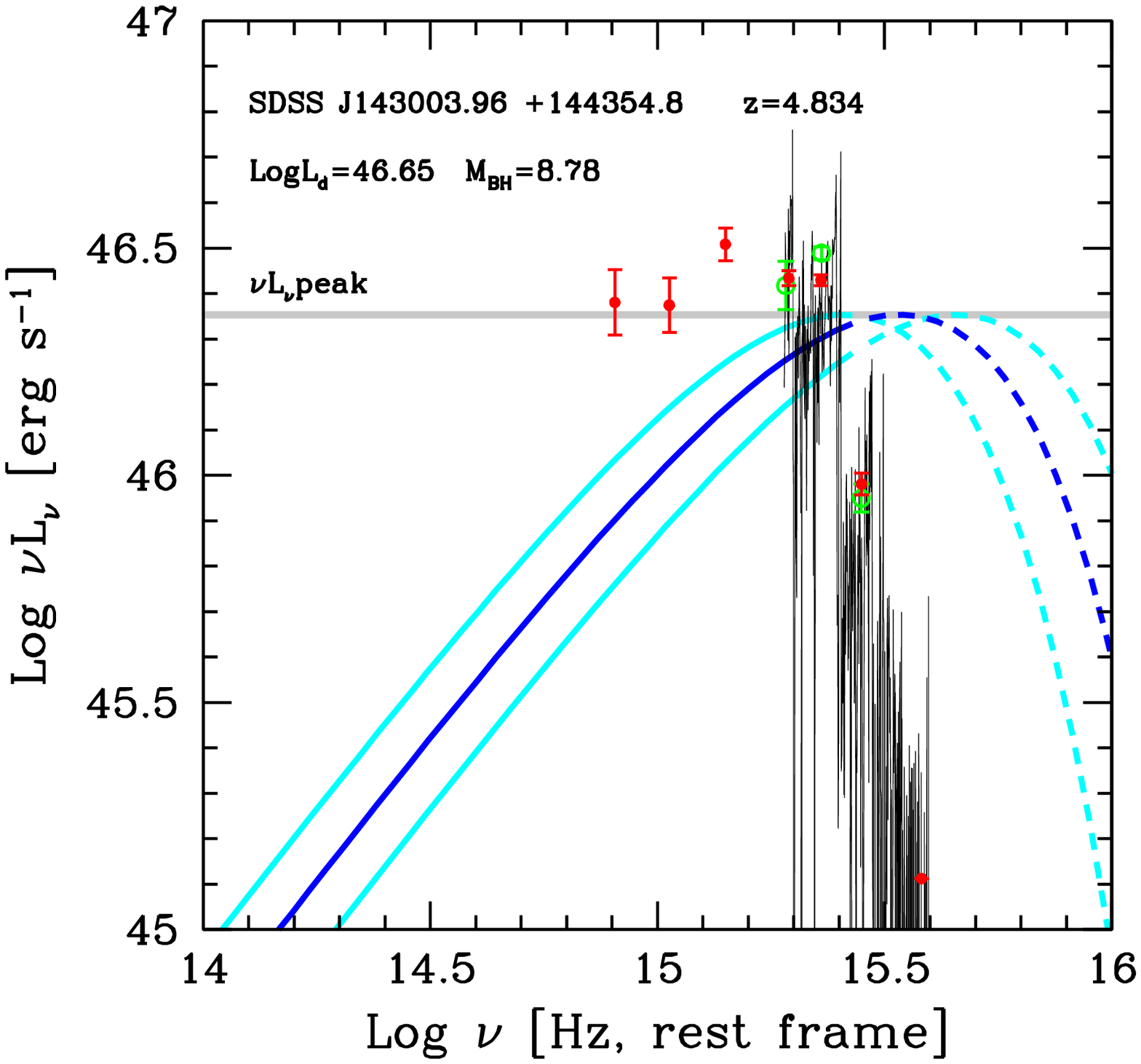,width=9cm,height=8cm}
\vskip -1.1 cm 
\hskip -0.4 cm
\psfig{figure=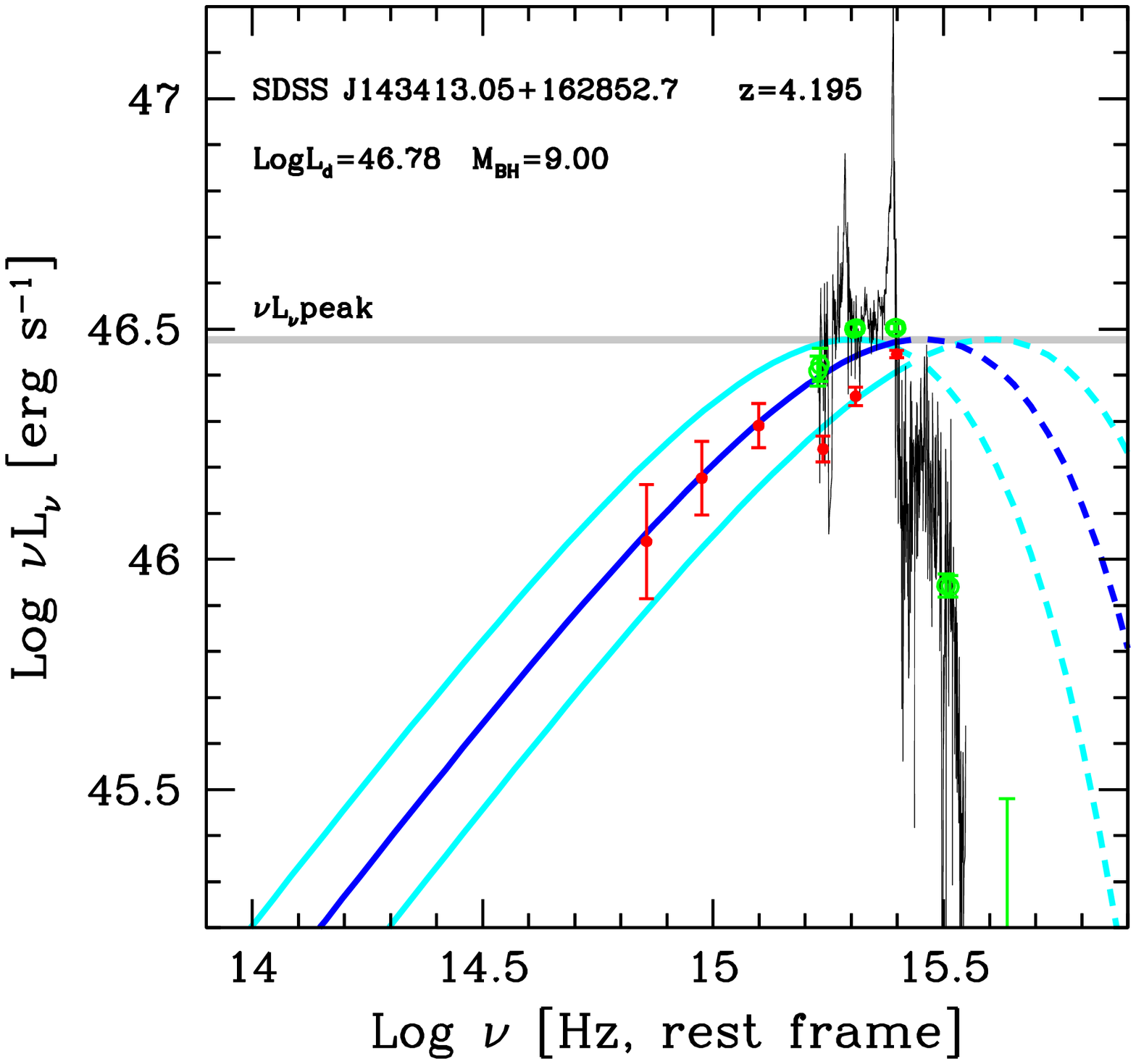,width=9cm,height=8cm}
\vskip -1.1 cm 
\hskip -0.4 cm
\psfig{figure=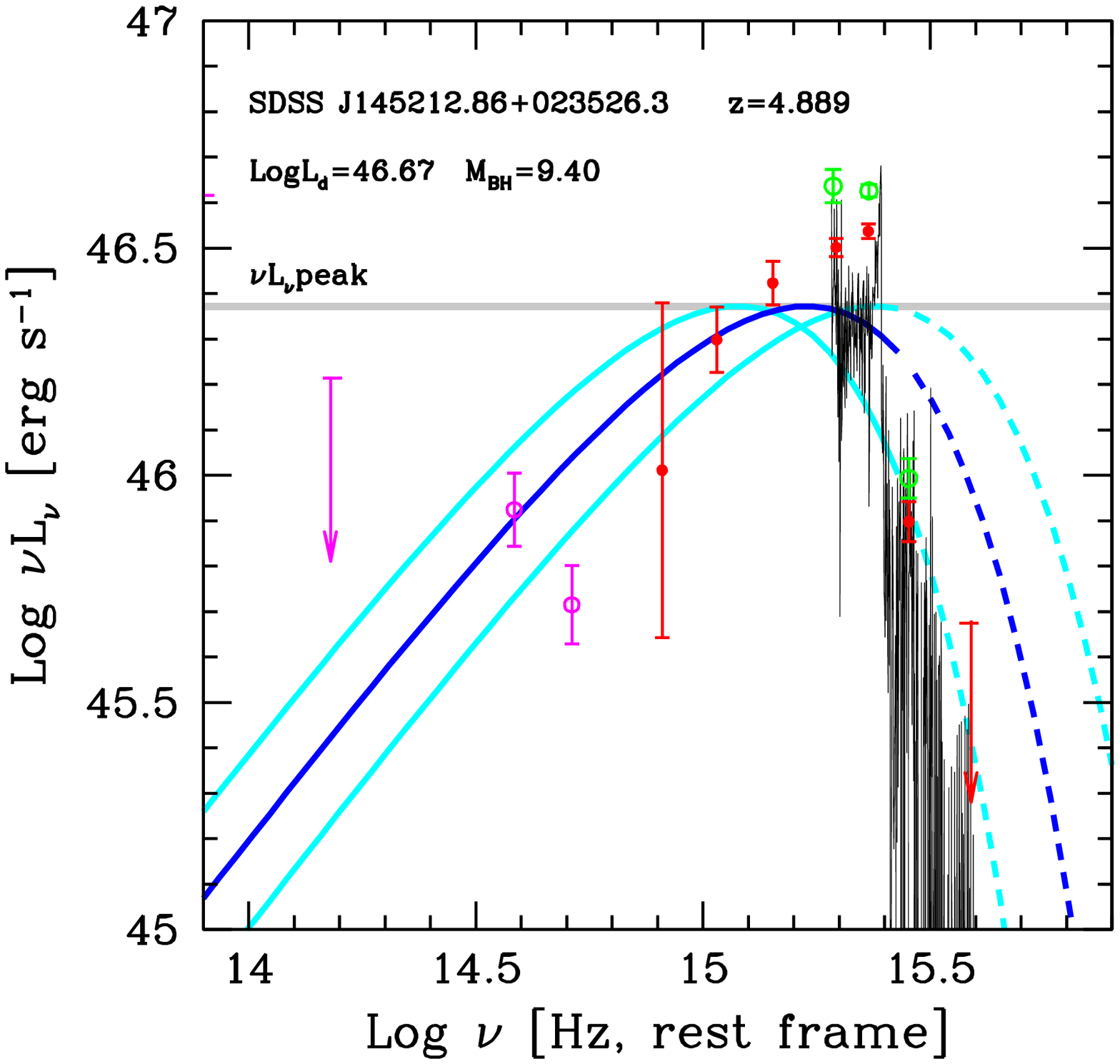,width=9cm,height=8cm}
\vskip -0.5 cm
\caption{
See caption of Fig.\ \ref{sed1}
}
\label{sed5}
\end{figure}

\begin{figure}
\vskip -0.6 cm 
\hskip -0.4 cm
\psfig{figure=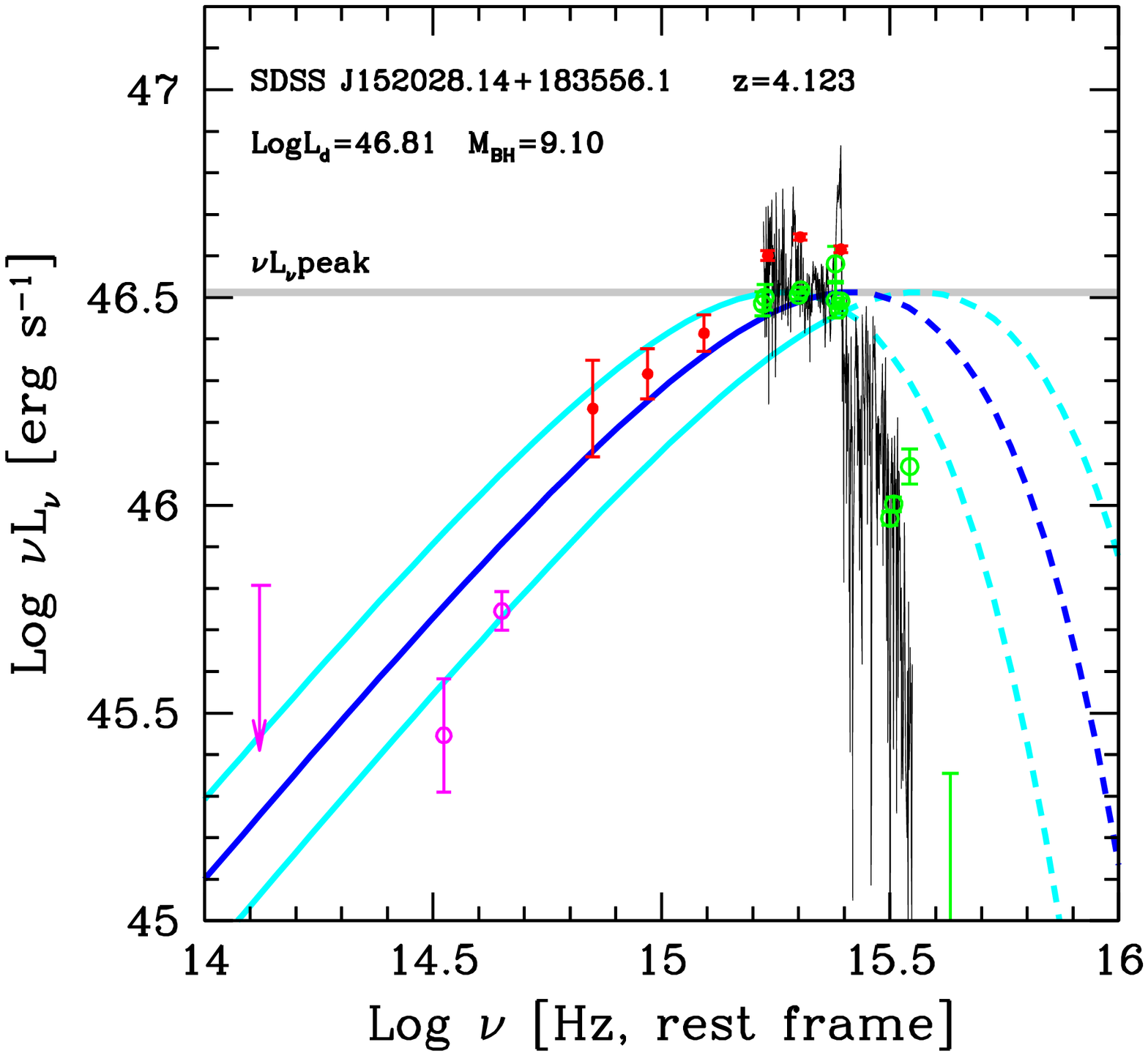,width=9cm,height=8cm}
\vskip -1.1 cm 
\hskip -0.4 cm
\psfig{figure=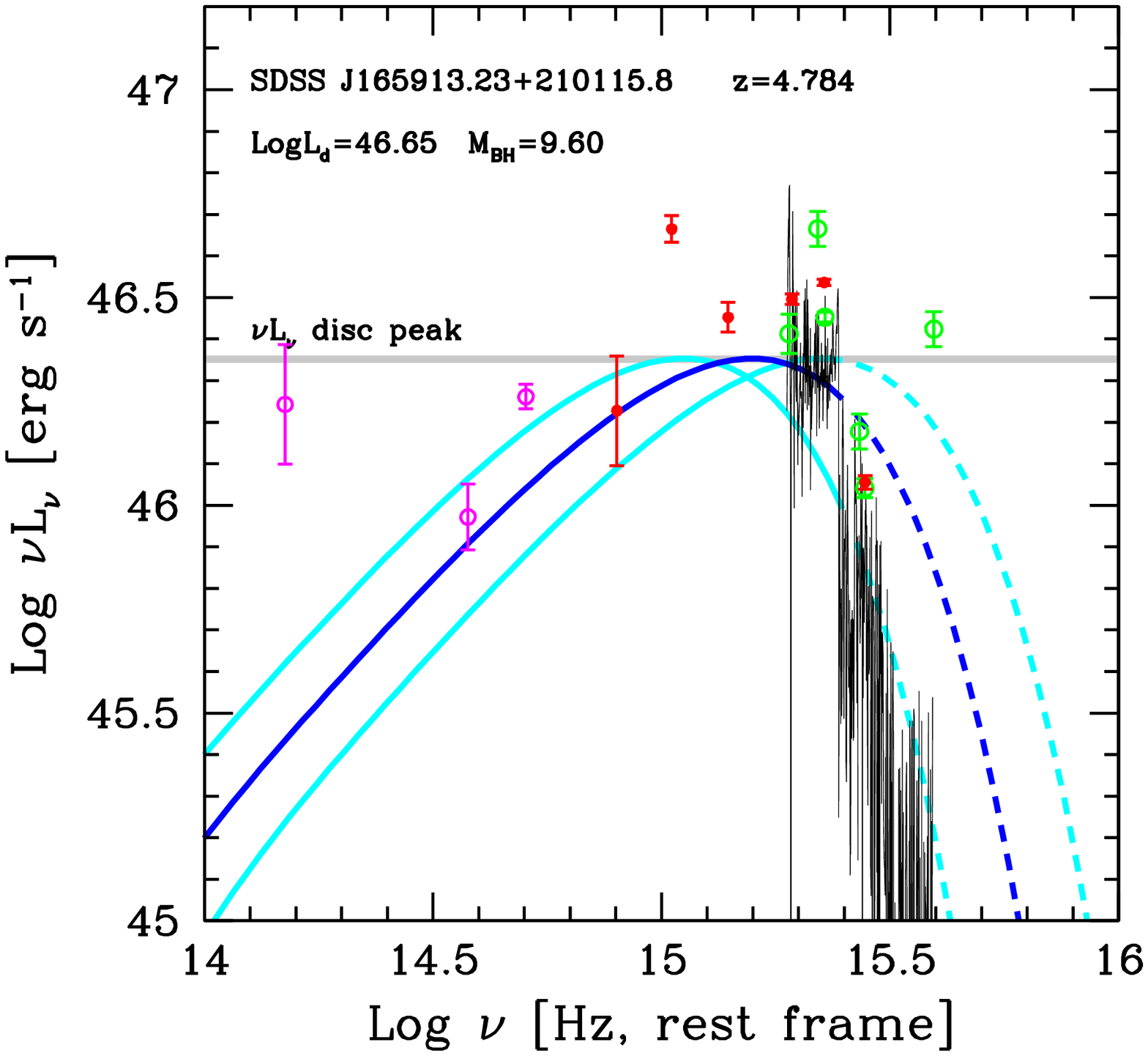,width=9cm,height=8cm}
\vskip -1.1 cm 
\hskip -0.4 cm
\psfig{figure=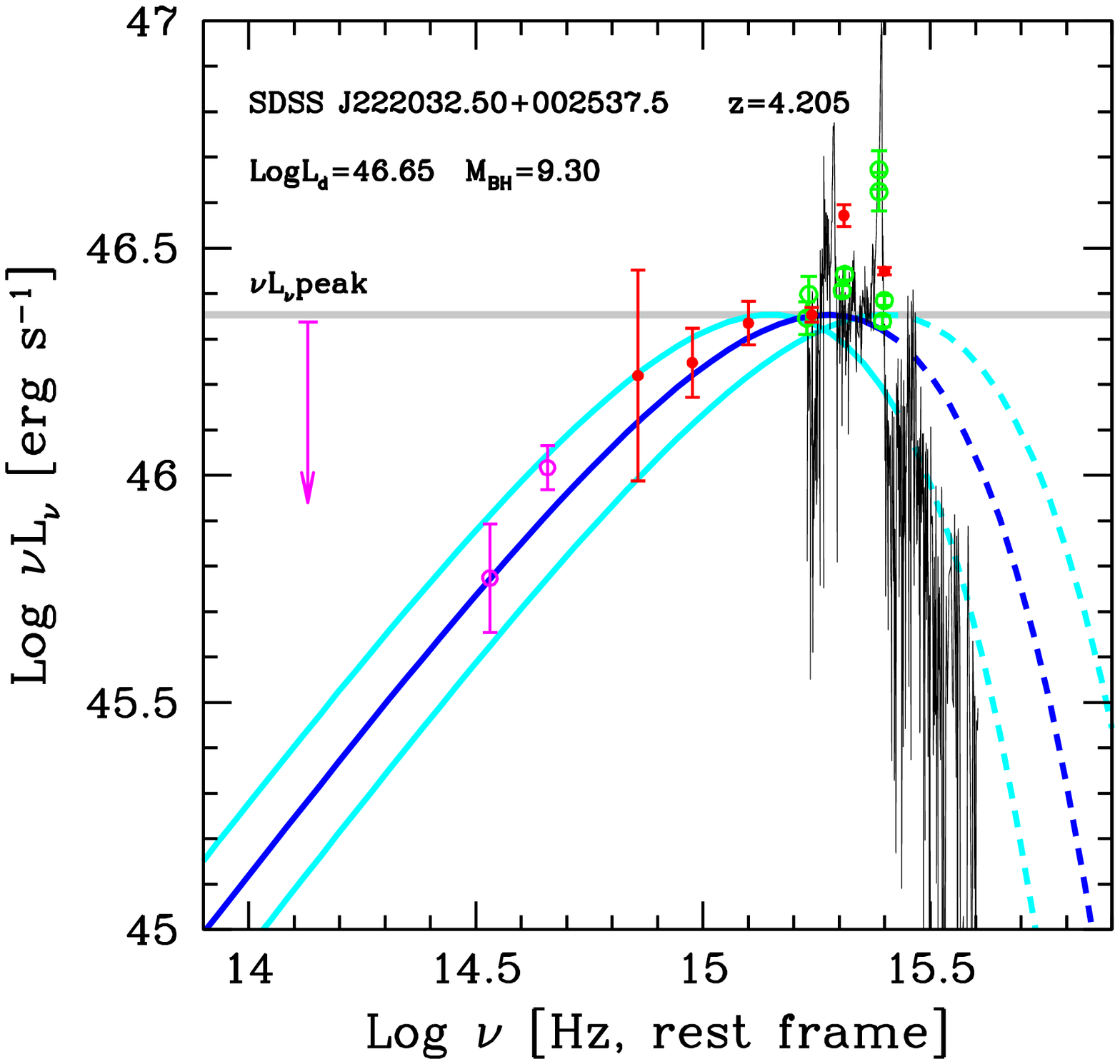,width=9cm,height=8cm}
\vskip -0.5 cm
\caption{
See caption of Fig.\ \ref{sed1}
}
\label{sed6}
\end{figure}

We have derived the black hole mass for all the sources in our sample.
Figs. \ref{sed1}--\ref{sed6} show the SEDs of our objects together with three models
with same disc luminosity but different masses. 
Note that the models do not fit the data at $\log\nu>15.4$, i.e.\ in the Ly$\alpha$ 
forest region, since the effect of the absorption produced by randomly distributed 
Ly$\alpha$ clouds along different lines of sight is unlikely to be predicted.
The lowest and highest masses indicate the
uncertainty associated to the measurement errors, possibe presence
of other contributions, and non strict simultaneity of the data.
The best values of the black hole masses we derived are listed in Tab. \ref{19}.
The indicated errors are not formal errors, but the mass range determined by 
applying the three models to the same data.

\begin{figure}
\vskip -0.6 cm 
\hskip -0.4 cm
\psfig{figure=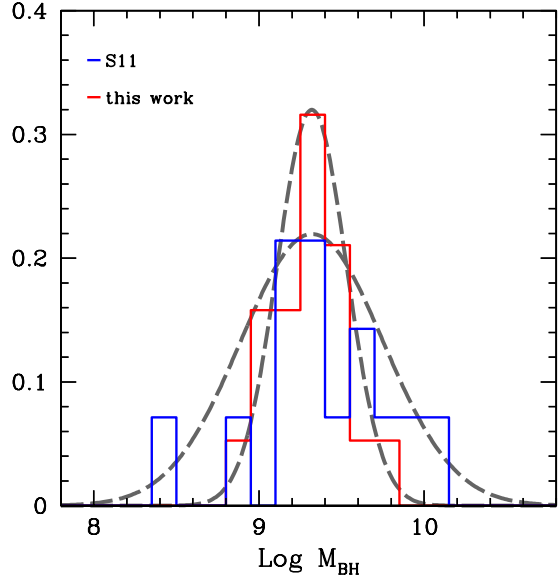,width=9cm,height=9cm}
\vskip -0.5 cm 
\caption{
Normalized distribution of the black hole masses of the quasars in our sample.
In blue, the histogram of the results obtained by S11 in their work.
In red, our results.
The dashed lines are best fit log--normal distributions, 
with values: 
$\langle\log M_{\rm BH}/M_\odot\rangle=9.37$; $\sigma=0.43$ (blue histogram, results of S11), 
$\langle\log M_{\rm BH}/M_\odot\rangle=9.31$; $\sigma=0.21$ (red histogram, this work). 
}
\label{histo}
\end{figure}

\begin{figure}
\vskip -0.6 cm 
\hskip -0.4 cm
\psfig{figure=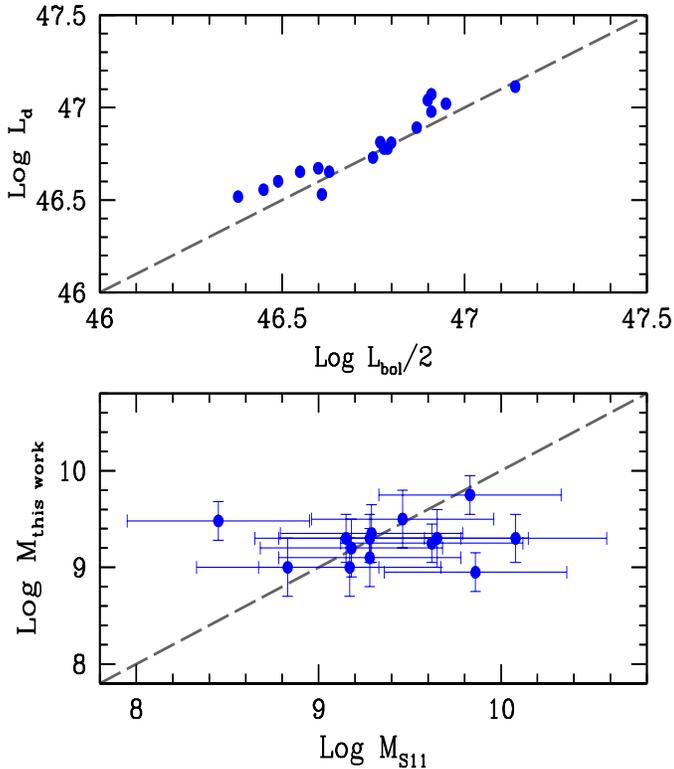,width=9cm,height=11cm}
\vskip -0.5 cm 
\caption{
Comparison of S11 with our results, including only the 14 sources out of our 
sample of 19 candidates with both the virial black hole mass and the 
accretion disc fitting estimates.
{\it Upper panel:}
the disc luminosity we obtain with the accretion disc fitting compared to 
half the bolometric luminosity (Calderone et al.\ 2013) as derived by S11.
SDSS J102623.61+254259.5 (B2 1023+25) is not included in this plot, 
since its redshift excludes from the SDSS wavelength window all the  
wavelengths from which the bolometric luminosity is typically derived 
(Richards et al.\ 2006).
{\it Lower panel:}
Values of the black hole masses as derived in this work with the accretion disc 
fitting (along with our error), compared with the virial masses derived by S11. 
This plot shows only the 14 objects with redshift that allows the virial black hole 
mass estimate.
The 5 candidates with the highest redshift, do not show the CIV broad line int 
the SDSS wavelength interval.
We assigned to the virial masses an average uncertainty of a factor 3.5 
(Vestergaard \& Peterson, 2006).
Note that the distribution of our $M_{\rm BH}$ estimates is definitely less dispersed 
than the distribution of the virial results (as can be clearly seen in Fig.\ \ref{histo}.
Our results show no systematics with respect to the S11 results.
}
\label{MLd}
\end{figure}

Since our sample is composed of highly luminous sources, given the survey flux limit
and the high redshift, we expect a large black hole mass, if the objects emit
at some large Eddington fraction, but do not exceed the Eddington limit.
This is indeed what we find:
all quasars in our sample host black holes with masses 
$M_{\rm BH}>10^9M_\odot$.
The red histogram in Fig. \ref{histo} shows the distribution of masses that we obtain 
for our sample using the fitting method.
Our results are distributed on a lognormal distribution, with an average value:
\begin{equation}
\Big\langle \log \frac{M_{\rm BH}}{M_\odot} \Big\rangle_{\rm this\, work} 
= 9.31 \pm 0.21 
\end{equation}
The blue histogram shown in Fig. \ref{histo} corresponds to the values of the 
black hole mass derived by S11 through the virial method (namely, through the
FWHM of the CIV line).
\begin{equation}
\Big\langle \log \frac{M_{\rm BH}}{M_\odot} \Big\rangle_{\rm S11} = 9.37 \pm 0.43 
\end{equation}
Comparing the two distributions, despite the paucity of objects, we 
note that the average value is the same, but the dispersion obtained
with the accretion disc fitting method is smaller.
The difference in dispersion is evident from the lower panel of Fig.\ \ref{MLd}, too.
This plot compares directly our $M_{\rm BH}$ estimates with the virial results 
(we include only the 14 sources for which S11 were able to virially derive the $M_{\rm BH}$).
The large dispersion of the S11 results prevents us from a clear ``source by source" 
comparison, but it is reasonable to assess that no systematic error seems to 
affect our estimates with respect to the virial black hole mass estimates.
Nevertheless, the great difference between the dispersions of the two methods 
is clearly an interesting issue, that 
will be investigated more deeply in the next section.

%----------------------------------------------------
%----------------------------------------------------
%----------------------------------------------------
%----------------------------------------------------

\section{Discussion}
\label{discussion}

In this work we constructed a sample of extremely radio--loud high--redshift quasars, 
in order to select a group of good blazar candidates.
We studied their thermal emission, deriving the luminosity emitted by the 
accretion disc (assumed as an optically thick, geometrically thin Shakura--Sunyaev disc) 
and the central black hole mass.
We took advantage of the good data coverage that can be achieved combining 
the WISE, SDSS and GROND data.
The optical--IR wavelength range, indeed, is where the Shakura--Sunyaev spectrum 
is expected to peak for SMBHs at $z>4$.
The peak of the emitted spectrum is indeed visible for our objects, 
and this greatly help to obtain good fits with a Shakura--Sunyaev disc model.

It is worth to notice that, 
once we assume an accretion disc model,  
the uncertainty of this method is strictly linked to the 
data uncertainties (i.e. the larger the data errors, the less constrained our estimate).
The wavelength coverage that we have in the IR--optical--UV band 
and the good precision of the available data, allow us to 
achieve an uncertainty of a factor of two.
Note that, in principle, we can obtain even more precise results with 
more accurate data.
In the virial method, instead, the precision of the mass estimates 
are driven by the uncertainties of the calibrating model,
that cannot be reduced below a factor 3--4 (Vestergaard \& Peterson, 2006).
Furthermore, the FWHM of the broad emission lines could be affected
by two systematic effects.
The first is related to the geometry of the BLR. 
If the clouds forming the BLR are distributed in a flattened 
geometry, their FWHM would be viewing angle dependent, being
smaller for small viewing angles (Decarli, Dotti \& Treves, 2011 ).
In turn, small viewing angles are expected for sources with a large radio--loudness.
Therefore, if such an effect plays a role, our sources would have virial black hole mass
estimates that are systematically smaller than what derived from the same sources
viewed at larger angles.
On the other hand, our accretion disc fitting is much less viewing angle dependent,
and thus the comparison between the two black hole mass distributions can tell
something about the BLR geometry.

The second effect concerns the effect of radiation pressure upon the line emitting
clouds, as discussed by Marconi et al.\ (2008, 2009).
Accounting for this extra force makes the virial estimate of the black hole mass
larger, and more so for near--Eddington disc luminosities.
Again, if the Shakura--Sunyaev disc is a reasonable description of the reality
even at luminosities near Eddington, then the estimate of the black hole mass
is independent of the disc to Eddington luminosity ratio $L_{\rm d}/L_{\rm Edd}$.

The comparison of the two black hole mass distributions gives these important
results:
\begin{enumerate}
\item
{\it The average black hole mass is the same ---}
As explained above, this suggests that the effect of both a flattened 
distribution of the clouds and/or the radiation pressure of the disc
radiation play a small role.
What inhibits a stronger conclusion is only the small number of objects,
but this issue surely deserves some attention in the future, 
when larger samples will be available.

\item
{\it The intrinsic width of the distribution must be extremely narrow ---}
%This favors our method in a direct comparison, since the errors on the estimates can be
%reduced with the achievement of better precision in the data.
%
As explained above, the virial method relies on calibrating some scaling
relation, as the size--luminosity relation, whose intrinsic scatter does
not allow a determination of the black hole mass more accurate than a factor 3 or 4.
The disc fitting method, instead, does not suffer from these limitations, but
depends only on the precision of the data.
For our sources, on average, we can assign an uncertainty of a factor 2.
Fig. \ref{histo} shows indeed that our results are less dispersed than 
the results obtained by S11.

% The availability of two different methods of measuring the mass with different precisions 
% give the possibility to investigate more deeply the intrinsic distribution of the black 
% hole masses.
% In principle, indeed, we could take advantage of having two different observed dispersions, 
% that come from the two different methods with which we measured the black hole masses 
% of the same sample.
Each observed dispersion
%dispersion that we obtain by measuring $M_{\rm BH}$ of a sample of objects
($\sigma_{\rm obs}$) 
is actually the convolution of the intrinsic dispersion of  the black hole masses 
($ \sigma_{M_{\rm BH}}$)
with the typical error that is done during each of the two measurements 
($\sigma_{\rm err}$):
\begin{equation}
\sigma^2_{\rm obs} = \sigma^2_{M_{\rm BH}} + \sigma^2_{\rm err}.
\end{equation}
Therefore, 
with two different observed distributions, we could be able to estimate the 
intrinsic dispersion of masses in our sample.

% Nevertheless, our sample is still too poor to draw firm statistical conclusion.
Bearing in mind that 
the sample is composed by only 19 objects, and that S11 obtained virial 
black hole masses for only 14  of these sources, let us compare  
% estimates of their black hole mass and disc luminosity.
% It is clear that the number of objects that would be necessary to obtain confidently 
% the intrinsic distribution of $M_{\rm BH}$ should be far larger.
% Taking in consideration 
the observed dispersions obtained by S11 
($\sigma_{\rm obs}^{\rm S11}=0.43$) with our dispersion
($\sigma_{\rm obs}^{\scriptscriptstyle\rm this\,work}=0.21$).
They are of the same order of the typical errors of the two 
methods ($\sigma_{\rm err}^{\rm S11}\sim0.4-0.5$ and 
$\sigma_{\rm err}^{\scriptscriptstyle\rm this\,work}\sim0.2-0.3$).
This suggests an extremely narrow intrinsic distribution, almost a $\delta$--function.
Interestingly, a very narrow $M_{\rm BH}$ distribution 
is also found by Calderone et al.\ (2013) for a sample 
of NLS1 galaxies, even if they consider a different range of masses and 
features of the sample.  
Nevertheless, this result has to be considered carefully.
First of all, we still have too few objects to draw statistically relevant conclusions.
Moreover, it could be affected by selection effects.
Indeed, our selection criteria collect the most luminous radio--loud quasars 
from the SDSS, that may already be composed by the ``tip of the iceberg" of the quasar 
population at high redshift. 
Since it is reasonable to expect that the most luminous are also the most massive quasars, 
our sample likely includes the objects with the most extreme black hole masses.
This possible selection effect may partially justify the narrowness of the intrinsic 
distribution, if we preferentially select the most luminous and massive objects.
Bearing in mind these observations, we can conclude that 
our small sample leads to the suggestion that the extremely radio--loud 
quasars located at $z>4$, and present in the SDSS, 
tend to have the same black hole mass, always larger than $10^9M_\odot$.

\end{enumerate}

%----------------------------------------------------
\begin{figure}
\vskip -0.6 cm 
\hskip -0.4 cm
\psfig{figure=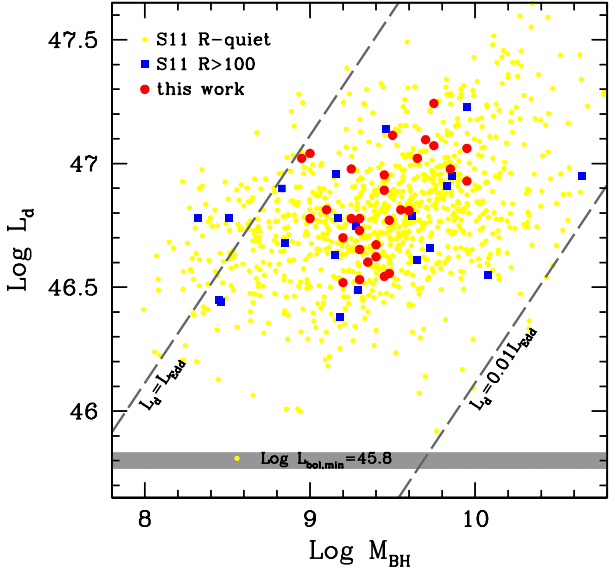,width=9cm,height=9cm}
\vskip -0.5 cm 
\caption{
Disc luminosity as a function of the black hole mass, both in logarithmic scale.
Blue squares are the results of S11 for the objects in our sample. 
Specifically, the $M_{\rm BH}$ is obtained with virial methods, 
while the $L_{\rm d}$ is half the bolometric luminosity derived 
as in Richards et al.\ (2006) (Calderone et al.\ 2013).
Red dots are the results of the method used in this work for the 19 objects in our sample.
Yellow dots are all the radio--quiet AGN included in S11, with $L_{\rm d}=L_{\rm bol}/2$.
The grey stripe highlights the minimum luminosity measured by S11 among the 
radio--quiet AGN.
The dashed lines indicate accretion regimes at the Eddington rate and at 
a hundredth of Eddington rate, as labelled.
Note that S11 obtain results that goes at disc luminosities larger than the Eddington limit.
}
\label{cfr}
\end{figure}
%----------------------------------------------------

We now turn our attention to the comparison between the disc properties of radio--loud
and radio--quiet AGN at $z>4$:
\begin{enumerate}
\item
{\it The thermal emission of radio--loud AGN do not differ from the analogous 
thermal emission of radio--quiet AGN ---}
Fig. \ref{cfr} shows the overall disc luminosity as a function of the black hole mass 
for the objects in our sample, as measured both by S11 (blue squares) 
and with our method (red dots), 
compared with all the radio--quiet AGN studied by S11 (yellow dots)
above $z=4$.
For the data obtained by S11, to estimate the disc luminosity 
we used half the bolometric luminosity 
calculated in their work, 
following Richards et al. (2006)
% through different bolometric corrections 
and Calderone et al. (2013).
Fig. \ref{cfr} shows that there are no intrinsic differences 
between radio--loud and radio--quiet AGN in their black hole mass and accretion luminosities.
% The radio--quiet S11 objects, indeed, have been selected with the same
% redshift limit as the extremely radio--loud ones that we analyzed, but they 
% have not a radio--counterpart according to the FIRST data.
% This ensures that we are trying to compare objects that differs only in their 
% radio--loudness (i.e.\ presence or absence of the jet).
% Fig. \ref{cfr} shows that 
Therefore the thermal emission from the accretion disc and 
the mass of the central black hole of AGN, in presence or absence of a jet, 
do not differ.
%This strengthens our assumption of similar details in the optical spectra 
%of radio--quiet and radio--loud AGN.
%Therefore, it is clear that looking for good blazar candidates among a sample of 
%generic quasars is an efficient approach.

\item
{\it Extremely radio--loud AGN at high redshift have high accretion rates ---}
Fig. \ref{cfr} illustrates also the range of accretion rates 
of the objects in our sample, in units of the Eddington one.
The two grey dashed lines in fact correspond to 
a Eddington rate ($L_{\rm d}=L_{\rm Edd}$) and a hundredth of Eddington 
($L_{\rm d}=10^{-2}L_{\rm Edd}$).
They are generally considered as the two physical limits of a radiatively efficient 
accretion operated by a Shakura--Sunyaev accretion disc.
Note that two extremely radio--loud objects, with the data obtained by S11, 
have a super--Eddington accretion rate.
On the contrary, our results never exceed this limit.
This suggests that at least in these cases our method is more reliable.
%
%{\it This is a proof of consistency of our model with respect to the virial methods 
%for measuring the black hole masses....}{\bf un po' troppo forte...}
%
It is also clear from Fig. \ref{cfr} 
that all our sources have high accretion rates.
Thus our sources, besides having a relatively large black hole mass,
[$\log(M_{\rm BH}/M_\odot)=9.3$], accrete at $\sim$10\% Eddington,
on average.
% This suggests that a high accretion rate could be linked to the high average 
% black hole mass of  our sample [$\log(M_{\rm BH}/M_\odot)=9.3$].
High rates of accretion justify the presence of many extremely massive 
black holes at high redshift, that perhaps are just ending the 
``fast growing age" through accretion at the Eddington rate.
Indeed, the formation and growth of the first supermassive
black hole is an open issue, that our results will hopefully help to solve.
%
%The formation of extremely massive black holes at high redshift is indeed 
%a still open problem.
%In order to deepen our knowledge about it, we require to have an
%estimate of how many extremely massive black holes are present at 
%high redshift.
% This is why we compile a coherent sample of blazar candidates at $z>4$.
%Blazars, indeed, thanks to their peculiar orientation, can be statistically 
%extremely relevant to compute the number of radio--loud, extremely massive, 
%high redshift AGN.

\item
{\it Blazar candidates ---}
% All sources in our sample have a black hole mass exceeding a billion solar masses.
The sample, selected through a cut in radio--loudness, preferentially collects
jets aligned with the line of sight. 
Among them, we expect to find a number of blazars, namely sources with 
a viewing angle comparable to or smaller than the beaming angle $1/\Gamma$.
If we knew the viewing angle of our sources, we could reconstruct the number density
of the entire (parent) population.
Specifically, 
if a source has a viewing angle smaller than the beaming angle, 
with an estimate of $\Gamma$ (through modelling and/or superluminal motion) 
we would be able to derive the total number of analogous objects, 
%One way to achieve this is to recognize, among our sample, which sources are
%blazars, namely oriented within a viewing angle smaller than $1/\Gamma$.
This is indeed the final goal of our research.
% The compilation and the study of this sample is oriented to select the best blazar 
% candidates, in order to observe them in the X--rays and confirm their blazar nature.
% The first result of our selection has been achieved by having already 
We have already identified as a blazar the more distant of the objects in our sample: B2 1023+25 
at $z\simeq 5.3$ (Sbarrato et al.\ 2012b).
In our previous work, we suggested a viewing angle smaller than $1/\Gamma$ 
for this source, as it is for the other two blazars of the original sample  
(SDSS J083946.22+511202.8 and SDSS J151002.92+570243.3, see \S \ref{sample}), 
Each of these blazars implies the existence of other $\sim 450(\Gamma/15)^2$ 
misaligned sources sharing the same characteristics, including the presence
of an actively accreting supermassive black hole, in the same region of the sky 
covered by the SDSS+FIRST survey.

\end{enumerate}

 \section{Conclusions}
 \label{concl}
 
 In this work, we composed a sample of 31 extremely radio--loud quasars at $z>4$. 
 We estimated the black hole masses of the 19 objects among them that are visible 
 from La Silla (Chile), and that are hence observable with GROND.
 The GROND data allowed us to observe directly the peak of the accretion disc 
 spectrum.
 Hence,  combining them with WISE and SDSS data, 
 we have been able to study the thermal emission of those objects, and to 
 derive with a precision of a factor of two their black hole masses.
 We found a very small range of masses for our sample, 
 that peaks at $M_{\rm BH} = 10^{9.3} M_\odot$.
 
 Our method, along with the good data coverage of the optical--IR band, 
 allowed us to derive both the black hole mass and the accretion disc luminosity 
 of the objects in our sample.  
 Hence we were able to estimate the accretion rate of the 19 extremely radio--loud quasars. 
 We found high accretion rate values for all of them.
 This could introduce interesting hints on the formation and (possibly fast) 
 growth of the first supermassive black holes at very high redshift.
 
 Our sample collects highly radio--loud quasars with extremely massive black holes.
 Such a radio--loudness criterion preferentially selects jets aligned near our line of sight.
 We expect to find in our sample a number of blazars, i.e.\ quasars seen with a 
 viewing angle comparable to (or smaller than) the jet beaming angle. 
 We will derive the jet orientation of the best candidates by fitting their broad--band SEDs.
 Specifically, we have planned for our three best candidates a set of 
 X--ray observations with the {\it Swift} X--ray Telescope. 
 X--ray data will provide hints on the jet non--thermal emission, helping us to constrain 
 the jet viewing angle. 
 
 Thanks to their peculiar orientation, 
 by identifying these objects as blazars we would be able to   
 derive the total number of analogous extremely massive 
 radio--loud objects with jet directed in other directions.
 In other words, with our approach 
 we expect to obtain the overall number of the parent population 
 of extremely massive blazars in the early Universe.
 This would allow a direct comparison of the density functions of 
 highly massive radio--quiet and 
 radio--loud objects even at very high redshift.

\section*{Acknowledgments}
We thank the anonymous referee for useful comments that helped improve the paper.
This research made use of the NASA/IPAC Extragalactic Database (NED) 
and part of it  is based on archival data, software or on--line services 
provided by the ASI Data Center (ASDC).
We made use of the data products from the Wide--field Infrared Survey Explorer, 
which is a joint project of the University
of California, Los Angeles, and the Jet Propulsion
Laboratory/California Institute of Technology, funded by the National
Aeronautics and Space Administration.
Part of the funding for GROND
(both hardware as well as personnel) was generously granted from the Leibniz
Prize to Prof. G. Hasinger (DFG grant HA 1850/28--1).

%----------------------------------------------------
%----------------------------------------------------

%----------------------------------------------------
%----------------------------------------------------

%----------------------------------------------------
%----------------------------------------------------

%----------------------------------------------------
%----------------------------------------------------

%----------------------------------------------------
%----------------------------------------------------

%----------------------------------------------------
%----------------------------------------------------

%------------------------------------------------
%------------------------------------------------
%------------------------------------------------

\end{document}